\documentclass[a4paper,11pt]{article}
\usepackage{jheppub} 
\usepackage{lineno}
\usepackage{float}
\usepackage[x11names,svgnames]{xcolor}

\title{Search for Light Dark Sectors Using Electron-Photon Collisions}

\author[a,b,c,d,1]{L. Angel\note{Corresponding author.}}
\author[e,f]{G. Casse}
\author[g]{G. Gambini}
\author[h]{A. S. de Jesus}
\author[i,j]{V. Kozhuharov}
\author[k]{A.A. Machado}
\author[a,b,l,m]{F. S. Queiroz}
\author[k]{E. Segreto}
\author[f]{J. D. Shergold}
\author[f]{J. Smirnov}

\affiliation[a]{Departamento de Física, Universidade Federal do Rio Grande do Norte, 59078-970, Natal, RN, Brasil}
\affiliation[b]{International Institute of Physics, Universidade Federal do Rio Grande do Norte, Campus Universitario, Lagoa Nova, Natal-RN 59078-970, Brazil}
\affiliation[c]{Department of Physics, University of Oslo, Box 1048, N-0316 Oslo, Norway}
\affiliation[d]{Max Planck Institut fur Kernphysik, Heidelberg, Germany}
\affiliation[e]{Oliver Lodge Laboratory, University of Liverpool, Liverpool L69 7ZE, United Kingdom}
\affiliation[f]{Department of Mathematical Sciences, University of Liverpool, Liverpool, L69 7ZL, United Kingdom}
\affiliation[g]{Facultad de Ciencias, Universidad Nacional de Ingeniería (UNI), Av. Túpac Amaru 210, Lima, Perú}
\affiliation[h]{Departamento de Física Matemática, Instituto de Física, Universidade de São Paulo, 05315-970 São Paulo, Brasil}
\affiliation[i]{Faculty of Physics, Sofia University, 5 J. Bourchier Blvd., 1164 Sofia, Bulgaria}
\affiliation[j]{INFN - LNF, Via E. Fermi 54 - 00044 Frascati, Italy}
\affiliation[k]{Instituto de Física ``Gleb Wataghin'', Universidade Estadual de Campinas (UNICAMP), Rua Sérgio Buarque de Holanda 777, CEP 13083-859, Campinas, São Paulo, Brazil}
\affiliation[l]{Millennium Institute for Subatomic Physics at the High-Energy Frontier (SAPHIR) of ANID, Fernández Concha 700, Santiago, Chile}
\affiliation[m]{Universidad de La Serena, Avenida Cisternas 1200, La Serena, Chile}

\emailAdd{lucia.correa.717@ufrn.edu.br}
\emailAdd{venelin.kozhuharov@cern.ch}
\emailAdd{farinaldo.queiroz@ufrn.edu.br}

\abstract{\noindent The dark photon is a new gauge boson that naturally arises in many beyond-the-Standard-Model scenarios and couples to the electromagnetic current through kinetic mixing. Owing to this feature, it is often considered a portal between the dark and visible sectors and has therefore become the target of numerous experimental searches worldwide.
In this work, we propose a search for dark photons based on inverse Compton scattering, $\gamma e^- \rightarrow A' e^-$, to be conducted at electron accelerators. In this setup, photons from a laser source impinge on an accelerated electron beam, producing a dark photon in the final state. We investigate two complementary search strategies based on the resulting missing-energy signature: a single-photon-veto approach and a measurement of the recoil-electron energy spectrum, with angular information considered as an extension of the latter. We derive the projected sensitivity assuming an incident photon energy of about 1~eV and an electron beam energy of 3~GeV. We show that this experimental setup could cover an unexplored region of parameter space and constitute a promising probe of dark sectors in the future.}

\begin{document}
\maketitle
\flushbottom

\section{Introduction}
\label{Intro}

\noindent The Standard Model (SM) of particle physics constitutes a successful theoretical framework that describes three of the four known interactions in nature: the strong, weak, and electromagnetic interactions. 
However, there are numerous phenomena for which the SM cannot provide a reasonable description. One of these phenomena is dark matter (DM), whose existence has been a subject of interest for many decades due to multiple pieces of evidence of its gravitational interaction, such as the observed versus expected matter discrepancy in the Coma Cluster, documented by Fritz Zwicky \cite{andernach2017englishspanishtranslationzwickys}; the rotation curve of galaxies, later observed by Vera Rubin and Kent Ford \cite{Vera}; the distribution of anisotropies in the cosmic microwave background (CMB) \cite{2020Planck}; the collision system of the Bullet Cluster \cite{Clowe:2006eq}; the formation of large-scale structures \cite{Blumenthal:1984bp}; and the amount of baryonic matter inferred by Big Bang nucleosynthesis (BBN) \cite{Copi:1994ev}, which is insufficient to explain the fluctuations in the CMB, among others.\\

The mere evidence of the existence of DM provides a strong indication that a dark sector beyond the known fundamental forces could exist \cite{Okun:1982xi,Holdom:1985ag,Preskill:1982cy,Burgess:2000yq}. Given that the SM can only account for a small fraction of the universe (approximately 5\% of it consisting of baryonic matter), it is not surprising to consider the possibility of the existence of a rich and still undiscovered dark sector filled with exotic particles, which are entitled to interesting signatures. 

This dark sector, often referred to as the ``hidden'' sector in the literature, could contain the so-called light and weakly coupled particles, which, as the name suggests, would interact very weakly with the known particles of the SM. In this project, the focus will be directed towards the study of the dark photon (DP), which is a sort of a massive photon with electromagnetic interactions similar to the photon. The presence of new vector field with such properties is considered to be the simplest extension beyond the SM. Due to its simplicity and resemblance to quantum electrodynamics, it has been the target of a multitude of experimental searches going from direct production at low-energy accelerators to astrophysical probes. In this work, we propose to take advantage of low-energy electron accelerators to access new physics at much lower energy scales using lasers. 


\noindent In this work, we will focus on the DP model. We propose here a collision between an electron beam and a photon from a laser, which could produce dark photons through dark inverse Compton scattering (DICS), $\gamma e^- \to A^\prime e^-$, analogous to standard inverse Compton scattering (ICS), covering a different mass range from previous dark photon searches in the past \cite{Chakrabarty:2019kdd,Su:2021jvk,Wang:2021xbi,Inan:2021dir,Oliveira:2022ypu}.


Generally speaking, this electron-photon collision can be applied to existing electron accelerators, but the experimental setup and physics scale involved depend very much on the technical details. In other words, the new physics reach can be determined once some accelerator parameters are laid out. That said, for concreteness, we will concentrate our discussion on 3~GeV electron beams existing at the Sirius accelerator \cite{Liu:2023xbm,Alves:2025czw}. 

This article is organized as follows: In Section~\ref{Model}, we present the main characteristics of the DP model, reviewing its theoretical features and framework. Section~\ref{Searches} covers the current status of experimental searches for dark photons across different contexts, ranging from helioscopes to light-shining-through-a-wall (LSW) experiments. In Section~\ref{DICS}, we describe the kinematics and theoretical details of the DICS process. Section~\ref{Proposal} is devoted to our experimental proposal, a novel setup to search for these particles. Here, we use the parameters of the Sirius facility for illustrative purposes, demonstrating the applicability of our setup in a realistic electron accelerator environment. In Section~\ref{Lower_Threshold}, we discuss the impact and benefits of low energy thresholds combined with wide angular acceptance. Section~\ref{Discussion} presents a detailed discussion on the expected SM backgrounds and the projected sensitivity of our setup using two complementary strategies exploiting the missing-energy signature: a photon-counting approach based on a single-photon veto, and a measurement of the recoil-electron energy spectrum, with angular information considered as an extension of the latter. Finally, our conclusions are given in Section~\ref{conclusions}.

\section{Dark Photon Model}
\label{Model}

The dark photon (DP), also known as the hidden photon or heavy photon and typically denoted as $A^\prime$, $\gamma^{\prime}$, or U, was first proposed by Bob Holdom \cite{HOLDOM1986196} in the context of a model featuring an additional renormalizable $U(1)_\chi$ local symmetry group 
\cite{FILIPPI2020100042}. In this framework, the dark photon mixes with the SM neutral gauge bosons through a kinetic mixing term, with the mixing represented by the parameter $\varepsilon$, which is usually considered to be small ($\varepsilon \ll 1$). This mixing establishes a portal between the SM and the dark sector, potentially allowing interactions between dark matter and visible matter \cite{Fabbrichesi_2021}. The dark photon could also be a viable candidate for bosonic dark matter, however, in this paper we only consider it as a mediator between DM and SM particles.\\


There are two types of dark photons: massive and massless, similar to the SM photon. However, the focus of the community is on the massive case, as its direct current-like coupling to ordinary matter could facilitate detection \cite{Fabbrichesi_2021,essig2013darksectorsnewlight}. Therefore, the free parameters of the model are the kinetic mixing $\varepsilon$ and the dark photon mass $m_{A^\prime}$.\\

The kinetic-mixing term between the SM photon and the vector boson of the new gauge symmetry is given by:
\begin{equation}
    \mathcal{L} = -\frac{\varepsilon}{2} F_{\mu \nu} \chi^{\mu \nu} \,,
\end{equation}
\noindent where $\varepsilon$ is the free parameter known as ``kinetic mixing'', $F_{\mu \nu}$ is the electromagnetic strength tensor, and $\chi_{\mu\nu}$ is the field-strength tensor for the new interaction mediated by the dark photon. This work assumes that the dark photon has mass, which can be acquired through the Stueckelberg mechanism \cite{Stueckelberg:1938zz, RUEGG_2004} or via spontaneous symmetry breaking of the gauge symmetry \cite{Nakayama_2021}. The kinetic terms are not diagonal in this basis, thus requiring a field redefinition to identify the physical gauge fields. After diagonalization, the Lagrangian that describes the interaction of the dark photon with the electromagnetic current of the SM can finally be expressed as:
\begin{equation}
    \mathcal{L} \supset -\frac{e \varepsilon}{\sqrt{1-\varepsilon^2}} J_\mu A^{\prime \mu} \simeq -e \varepsilon J_\mu A^{\prime \mu} \,.
\end{equation}

\section{Current Status of Experimental Searches}
\label{Searches}

\begin{figure*}[ht!]
    \centering
    \includegraphics[scale=0.58]{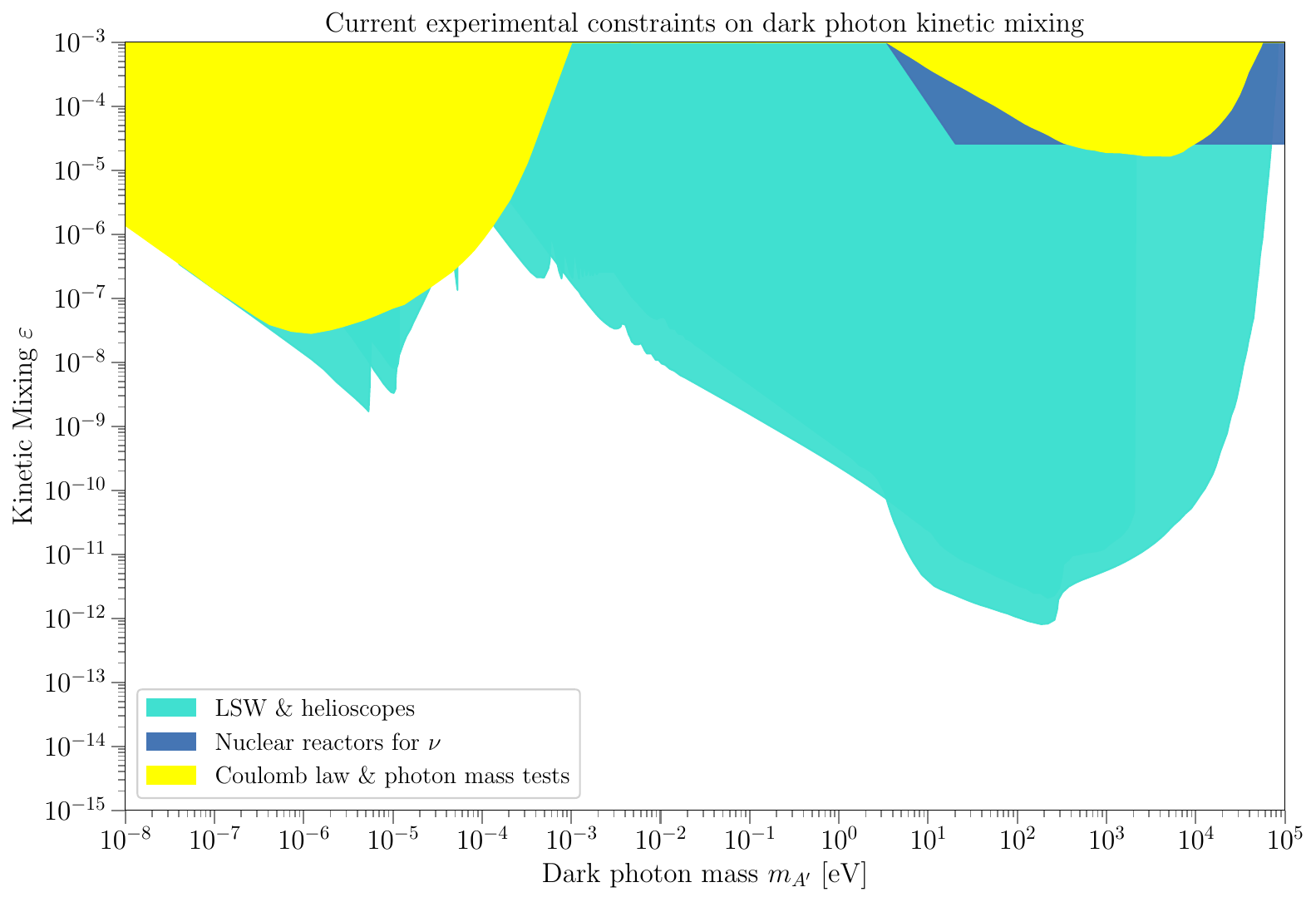}
\caption{Upper exclusion limits in the parameter space of the dark photon mass $m_{A^{\prime}}$ (x-axis) and the kinetic mixing parameter $\varepsilon$ (y-axis). This figure shows only current experimental bounds, excluding those that assume the dark photon is a dark matter candidate, which is not considered in this work. Data taken from \cite{AxionLimits}}
    \label{fig:Upper_Limits}
\end{figure*}

Since the DP couples to the electromagnetic current in a manner analogous to the SM photon, its production mechanisms mirror those of ordinary photons. The primary production channels include Bremsstrahlung ($e^{-}Z \rightarrow e^{-}Z A^{\prime}$), annihilation ($e^{-}e^{+}\xrightarrow{}\gamma A^{\prime}$), Drell-Yan ($q\Bar{q}\xrightarrow{}A^{\prime}\xrightarrow{}l^{+}l^{-}$ or $h^{+}h^{-}$), meson decays ($M \xrightarrow{}\gamma A^{\prime}$) and \textcolor{violet}{DICS} ($e^{-}\gamma \xrightarrow{}e^{-}A^{\prime}$), which constitutes the main focus of this work.\\

Based on the different production mechanisms, various experiments have been developed using different techniques, primarily centered on the decay modes (visible or invisible), depending on the mass range and experimental properties. 
The following will briefly outline each of the sources of the current experimental constraints on the mass of the dark photon and the kinetic mixing parameter $\varepsilon$ for $m_{A^{\prime}} <1$ MeV, which is our region of interest: \\

\begin{itemize} 

\item \textit{LSW \& Helioscopes:}
LSW is an experimental method used in the search for Beyond the Standard Model (BSM) particles, particularly Weakly Interacting Sub-eV Particles (WISPs), taking advantage of their possible mixing with the SM photon \cite{Redondo_2011}. This method was first proposed in 1982 \cite{Okun:1982xi} to search for dark photons. It was later adapted to explore axion-like particles \cite{Anselm:1985obz,VanBibber:1987rq}.
The method involves directing a high-power laser through a region with a strong magnetic field. The interaction between the photons and the magnetic field can generate a beam of new particles. Unlike the original laser beam, these particles may be capable of passing through a solid barrier (e.g., a wall). After passing through the barrier, a second magnetic field of comparable strength redirects the newly produced particles toward a detector \cite{Redondo_2011,UBDM}.
In the case of helioscopes, their setup consists of a long, sealed cavity that points toward the Sun. Inside this cavity, a powerful magnetic field is generated. At the end of the cavity, there is an X-ray detector sensitive to weak signals. Originally, the magnetic field was introduced to facilitate the oscillation of axions into photons. However, the setup can be adapted for the case of the dark photon, where the magnetic field is not necessary, as these particles can naturally mix with the SM photon. The probability of a dark photon converting into a photon depends on its mass and the length of the cavity  \cite{Redondo_2008}. 

These experiments provide an ideal framework for exploring BSM physics. The experimental constraints of both types of experiments 
can be directly applied to the case of massive dark photons. Among the main experiments of this type are: CAST \cite{Redondo_2008}, ALPS \cite{ALPS}, CROWS \cite{CROWS}, DARK SRF \cite{SRF}, HINODE \cite{HINODE}, IAXO \cite{IAXO}, LSW-AMDX \cite{PhysRevD.88.112004}, LSW-CERN \cite{betz2012microwaveparaphotonaxiondetection}, LSW-Spring-8 \cite{INADA2013301}, LSW-UWA \cite{PhysRevD.82.052003}, and SHIPS \cite{Schwarz_2015}. These experiments continue to play a critical role in constraining the properties of dark photons and other BSM particles. The bounds from these experiments are presented in {\it turquoise} in Figure~\ref{fig:Upper_Limits}. \\

\item \textit{Nuclear Reactor Experiments:} Nuclear reactor-based experiments, such as GEMMA or TEXONO (Taiwan EXperiment On Neutrino), were originally designed to study processes like $e^{-}\nu_{e^{-}}\xrightarrow{} e^{-}\nu_{e^{-}}$. However, their results can also be interpreted in the context of photon-dark photon oscillations inside the reactor, much like the well-known neutrino oscillations. These experiments provide an additional avenue for studying dark photons due to the high intensity of electromagnetic fields present in nuclear reactors. The results published in \cite{TEXONO} conclude that neutrino experiments are insensitive to dark photons with masses below $\sim 0.1$ eV. The bounds from Nuclear Reactors are presented in {\it blue} in Figure~\ref{fig:Upper_Limits}. 

\item \textit{Atomic and Coulomb law experiments:} The possible existence of a dark photon and its kinetic mixing parameter with ordinary photons could modify the Coulomb force at atomic distances, introducing an additional Yukawa-type potential,
\begin{equation*}
V(r) = -\frac{Z \alpha}{r}\left(1+e^{-m_{A^{\prime}} r} \varepsilon^2\right) \,,
\end{equation*}
where $m_{A^{\prime}}$ is the mass of dark photon and $\frac{Z \alpha}{r}$ is the usual Coulomb potential.
The inclusion of this potential would perturb atomic energy levels, thereby enabling the indirect inference of a dark photon through high-precision spectroscopy. These modifications to the energy levels would alter the Rydberg limit established for the pure Coulomb case and would be observable in spectral transitions. Corrections to the Rydberg constant and the Lamb shift could experimentally translate into new limits on the parameter space of the dark photon, specifically its kinetic mixing term and its mass \cite{PhysRevD.2.483,Jaeckel_2010}. Key experiments designed to probe these effects include Plimpton-Lawton experiment, Cavendish-like experiments, AFM, Atomic Spectroscopy, among others \cite{Kroff_2020}. The bounds from Atomic and Coulomb law experiments are presented in {\it yellow} in Figure~\ref{fig:Upper_Limits}.

\end{itemize} 

In this work, however, we consider two complementary search strategies exploiting the missing-energy signature associated with DICS. The first is based on the photon-counting technique and a single-photon veto, in which the absence of the SM photon expected from an ICS event is used to identify a potential dark-photon signal. Photon counting, as the name suggests, is a technique that involves counting individual photons using a photon counting detector (PCD). When a photon interacts with the detector's material, it generates an electrical pulse whose amplitude is proportional to the photon's energy. Among its main advantages, compared to traditional detectors, is the reduction of electronic noise, since measuring each photon individually results in a clearer signal. Additionally, this technique allows one to determine the energy of each photon through the use of energy thresholds. Some of the main detectors used in this method include photomultipliers, Geiger counters, and scintillation counters \cite{2009NaPho...3..696H,Photo-counting,KREISLER2022110229}. This technique is especially useful when dealing with processes with photons in the final state, being capable to measure low-energy photons. The second strategy is based on the measurement of the recoil-electron energy spectrum, with angular information considered as an extension of this analysis.

\section{The Dark Inverse Compton Scattering}\label{DICS}

\begin{figure}[htb]
  \centering
  \begin{minipage}{0.48\linewidth}
    \centering
    \includegraphics[width=\linewidth]{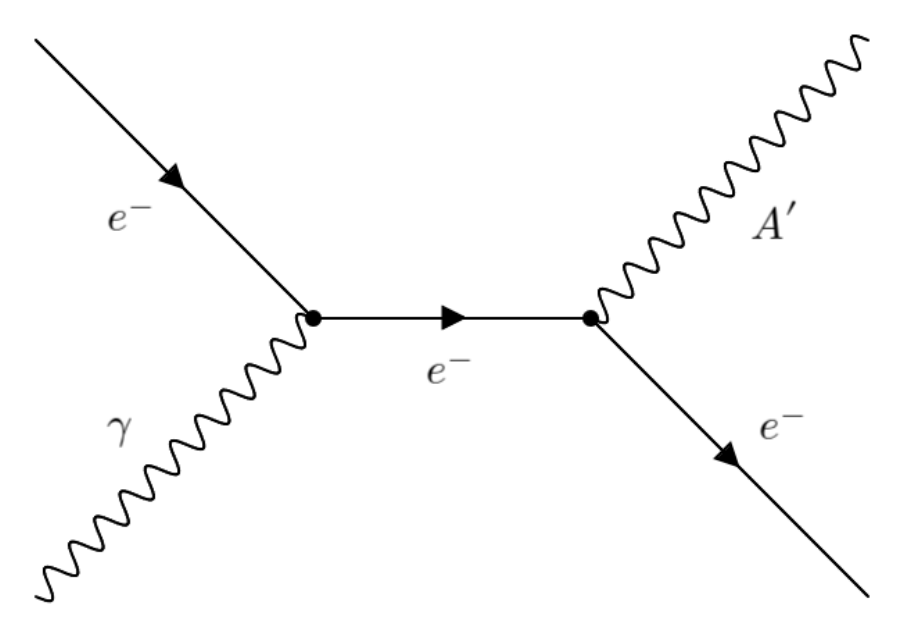}
    \\[1ex]
    {\small (a) s-channel}
  \end{minipage}\hfill
  \begin{minipage}{0.46\linewidth}
    \centering
    \includegraphics[width=\linewidth]{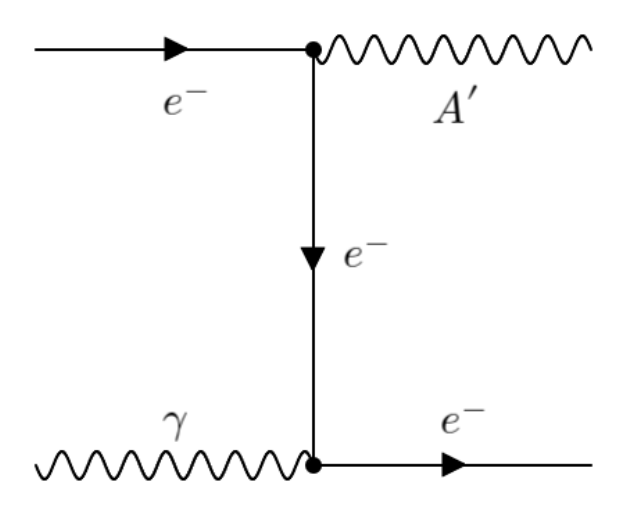}
    \\[1ex]
    {\small (b) t-channel}
  \end{minipage}
  \caption{Feynman diagrams contributing to DICS.}
  \label{fig:feynman_diagrams}
\end{figure}

\noindent The search for dark photons via the kinetic-mixing portal can be effectively pursued through the DICS process, $e^-\gamma \rightarrow e^-A'$. This mechanism, represented by the $s$- and $t$-channel Feynman diagrams in Figure~\ref{fig:feynman_diagrams}, is analogous to ICS, with the key difference that the outgoing photon is replaced by a dark photon. The DICS cross section is proportional to the ICS one, scaled by the kinetic-mixing parameter squared, i.e. $\sigma_{\mathrm{DICS}} \approx \varepsilon^2 \sigma_{\mathrm{ICS}}$. 
The full expression, which depends on the dark-photon mass ($m_{A'}$) and the center-of-mass energy ($\sqrt{s}$), is given in Eq. \ref{eq:cross}.

\begin{figure*}[ht!]
\begin{equation}
\label{eq:cross}
\begin{aligned}
    \sigma &= \frac{g^2 {g'}^2}{8 \pi (s-m_e^2)^3} \Bigg[ 
    \frac{ s^3+15 m_e^2 s^2-m_e^4 s+ m_{A'}^2 \left(7 s^2+2 m_e^2 s-m_e^4\right) +m_e^6 }{2 s^2} \\
    &\qquad \times \sqrt{\left(s-m_e^2\right)^2-2 m_{A'}^2 (s+m_e^2)+m_{A'}^4} + (s^2-6m_e^2s-2m_{A'}^2(s-m_e^2)-3m_e^4+2m_{A'}^4) \\
    &\qquad \times \log \left(\frac{s+m_e^2-m_{A'}^2+\sqrt{\left(s-m_e^2\right)^2-2 m_{A^{\prime}}^2 (s+m_e^2)+m_{A'}^4}}{s+m_e^2-m_{A'}^2-\sqrt{\left(s-m_e^2\right)^2-2 m_{A'}^2 (s+m_e^2)+m_{A'}^4}}\right) \Bigg] \,.
\end{aligned}
\end{equation}
\hrulefill
\end{figure*}

\noindent Here, $g=e$ is the ordinary electron-photon QED coupling, and $g'\simeq e\,\varepsilon$ is the effective electron-dark-photon coupling induced by kinetic mixing. With this identification, the prefactor in Eq.~\ref{eq:cross} satisfies $g^2{g'}^2\simeq e^4\varepsilon^2$. In the limit $m_{A'}\to0$ and $g'\to g$, the ordinary ICS result is recovered.\\

Despite the process being general to electron-photon collisions, the present analysis is inspired by the existing infrastructure at the Brazilian Synchrotron Light Laboratory (LNLS), which hosts an electron accelerator named SIRIUS. This is needed because the experimental setup and new physics reach heavily rely on the technical details of the accelerator setup.

\begin{figure*}[ht!]
    \centering
    \includegraphics[scale= 0.6]{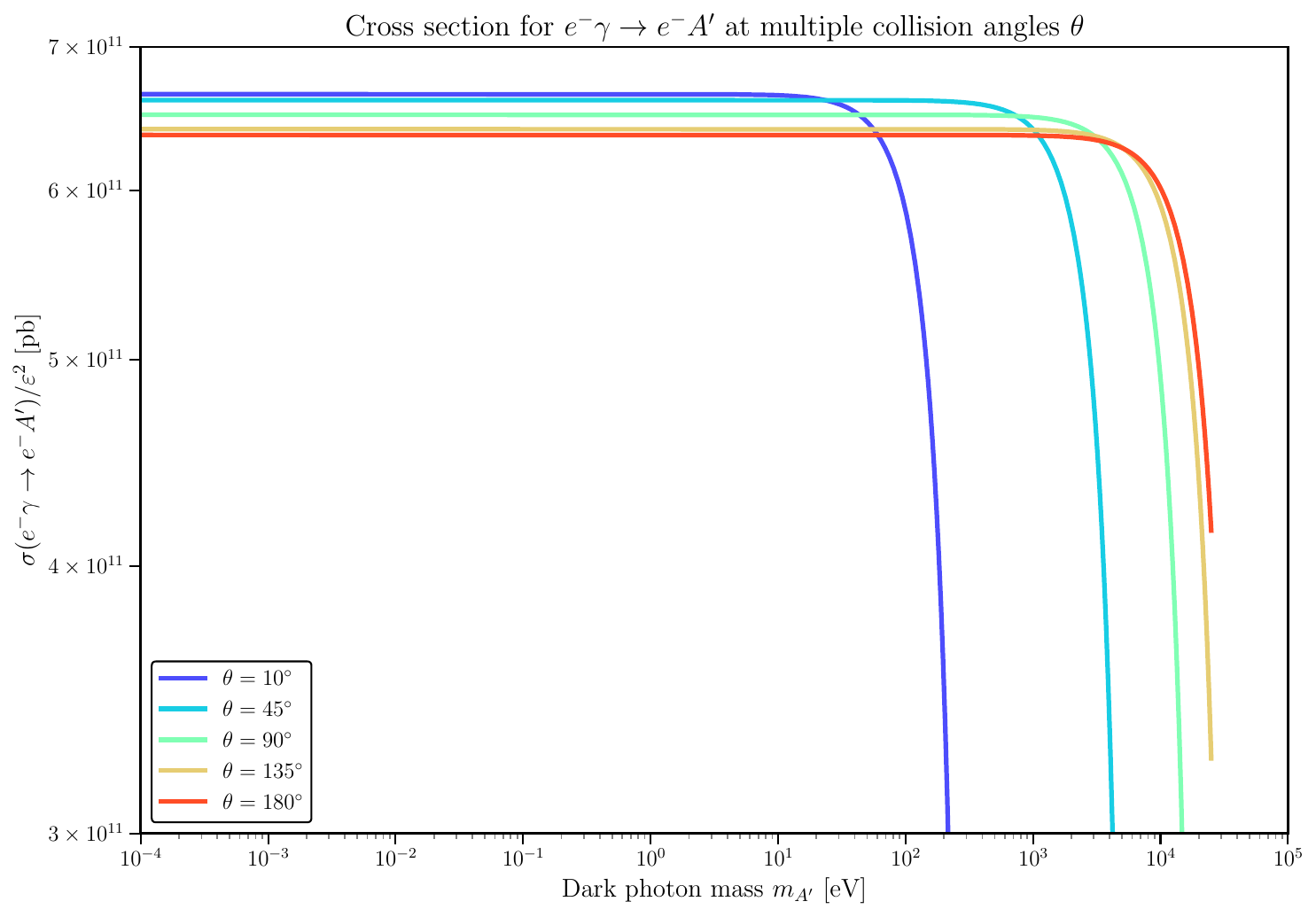}
    \caption{Cross section for $e^-\gamma \to e^- A'$ as a function of the dark photon mass $m_{A'}$ for different collision angles $\theta$. The plotted cross section is normalized by $\varepsilon^2$, allowing the results to be rescaled for any choice of the kinetic mixing parameter $\varepsilon$.}
    \label{fig:cs_mDP}
\end{figure*}

The cross-section behavior as a function of $m_{A'}$ (Figure~\ref{fig:cs_mDP}), evaluated for various collision angles $\theta$, shows that the ideal head-on configuration ($\theta = 180^\circ$) maximizes sensitivity to higher dark-photon masses due to the angular dependence of $\sqrt{s}$. Differences between $\theta = 135^\circ$ and $\theta = 180^\circ$ are minimal, which justifies the head-on configuration adopted in our projections and enables sensitivity to masses up to $\sim 1.2\times 10^4$~eV. At very small angles ($\theta < 10^\circ$), the cross section develops infrared divergences that are rendered irrelevant by the experimental energy thresholds.

\begin{figure*}[ht!]
    \centering
    \includegraphics[scale=0.47]{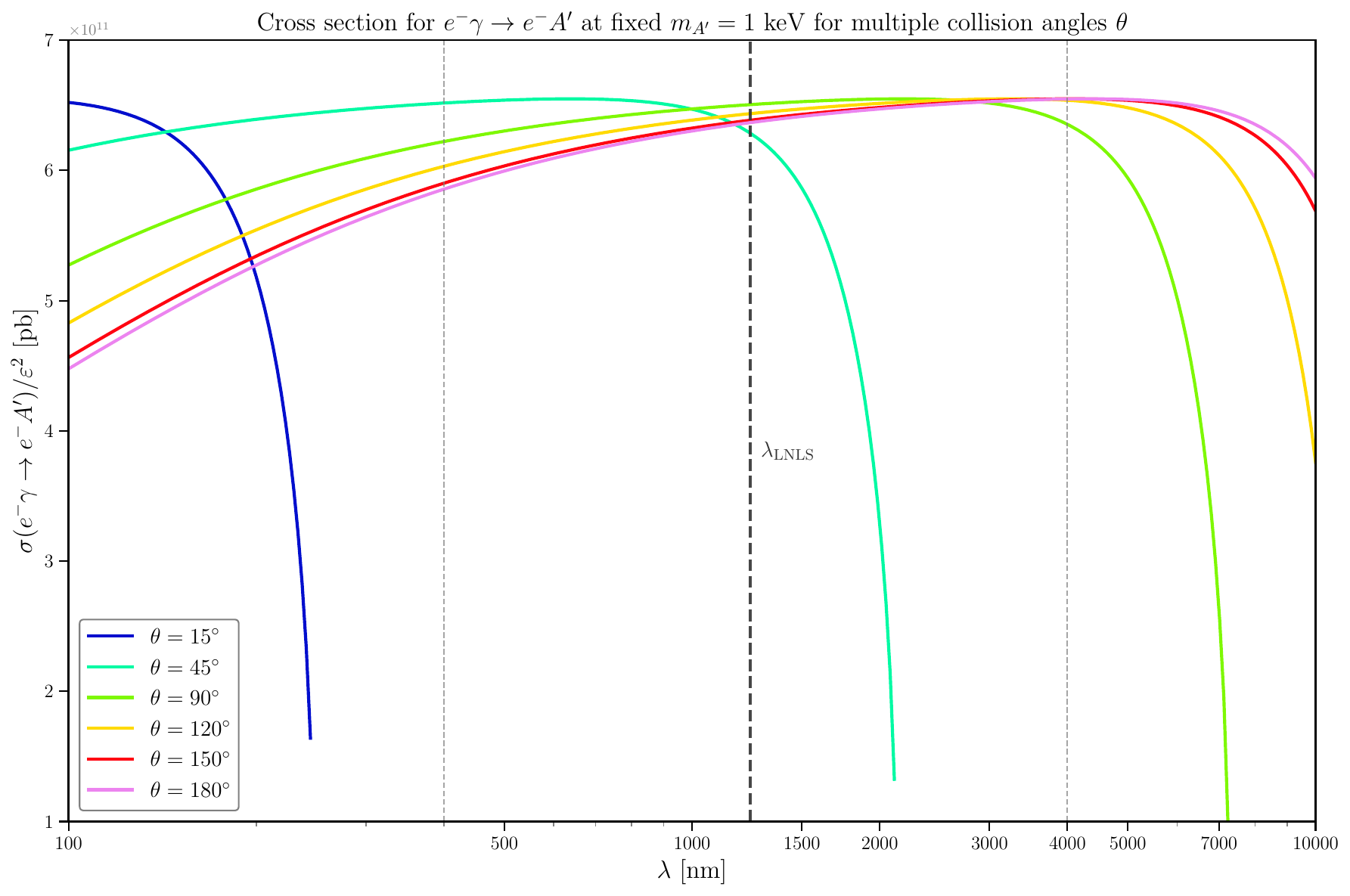}
    \caption{Cross section $\sigma(e^-\gamma \to e^- A')$ normalized by $\varepsilon^2$ as a function of the laser wavelength $\lambda$ for different collision angles $\theta$, with fixed dark photon mass $m_{A'} = 1~\mathrm{keV}$. The vertical gray dashed lines indicate the optimal sensitivity region, $400~\mathrm{nm} \leq \lambda \leq 4000~\mathrm{nm}$, and the black dashed line shows the wavelength used in this work ($\lambda \simeq 1240~\mathrm{nm}$), corresponding to a 1~eV laser photon. This figure is shown for illustrative purposes to highlight the variation of the cross section with wavelength and incidence angle.}
    \label{fig:angles}
\end{figure*}

\noindent For numerical estimates, we employ the parameters of the Sirius accelerator: an electron-beam energy of 3 GeV and a laser-photon energy of 1 eV ($\lambda = 1240$ nm). The dependence of the cross section on the laser wavelength (Figure~\ref{fig:angles}), shown here for $m_{A'} = 1$~keV, indicates an optimal sensitivity band for $\lambda$ between $400$~nm and $4000$~nm. The chosen 1~eV laser lies well within this region, particularly for $\theta \geq 135^\circ$. For more oblique collisions ($\theta < 45^\circ$), shorter-wavelength lasers would be advantageous to compensate for the reduced phase space.

\noindent Since the dark photon is expected to be invisible in the sub-MeV mass range considered here, detection must rely on indirect methods. We consider two complementary strategies exploiting the resulting missing-energy signature. The first combines photon counting with a single-photon veto, using the absence of the photon expected from the corresponding ICS process to identify a potential DICS event. The second is based on the recoil-electron energy spectrum, with angular information considered as an extension of this analysis. The following section presents the proposed experimental setup and its implementation at an electron accelerator.

\section{Experimental Setup}
\label{Proposal}

\noindent This work presents a preliminary sensitivity study and a proposal for searching for low-mass dark photons. As discussed in the previous section, this study could be generalized to different electron accelerators that could be operated to conduct electron-photon collisions. However, the proposal of an experimental setup demands more technical details, and for this reason, we will focus our analysis on the existing electron accelerator at Sirius \cite{Alves:2025czw}. The facility accelerates electrons to $3~\mathrm{GeV}$ and provides high-brilliance synchrotron radiation, which is used in many fields of science, such as biology, chemistry and others. Our focus, however, is on the electron beam, which can be used to investigate BSM physics. Particularly, we propose to search for DP production through DICS.

\begin{figure*}[htbp!]
    \centering
    \includegraphics[scale=0.5]{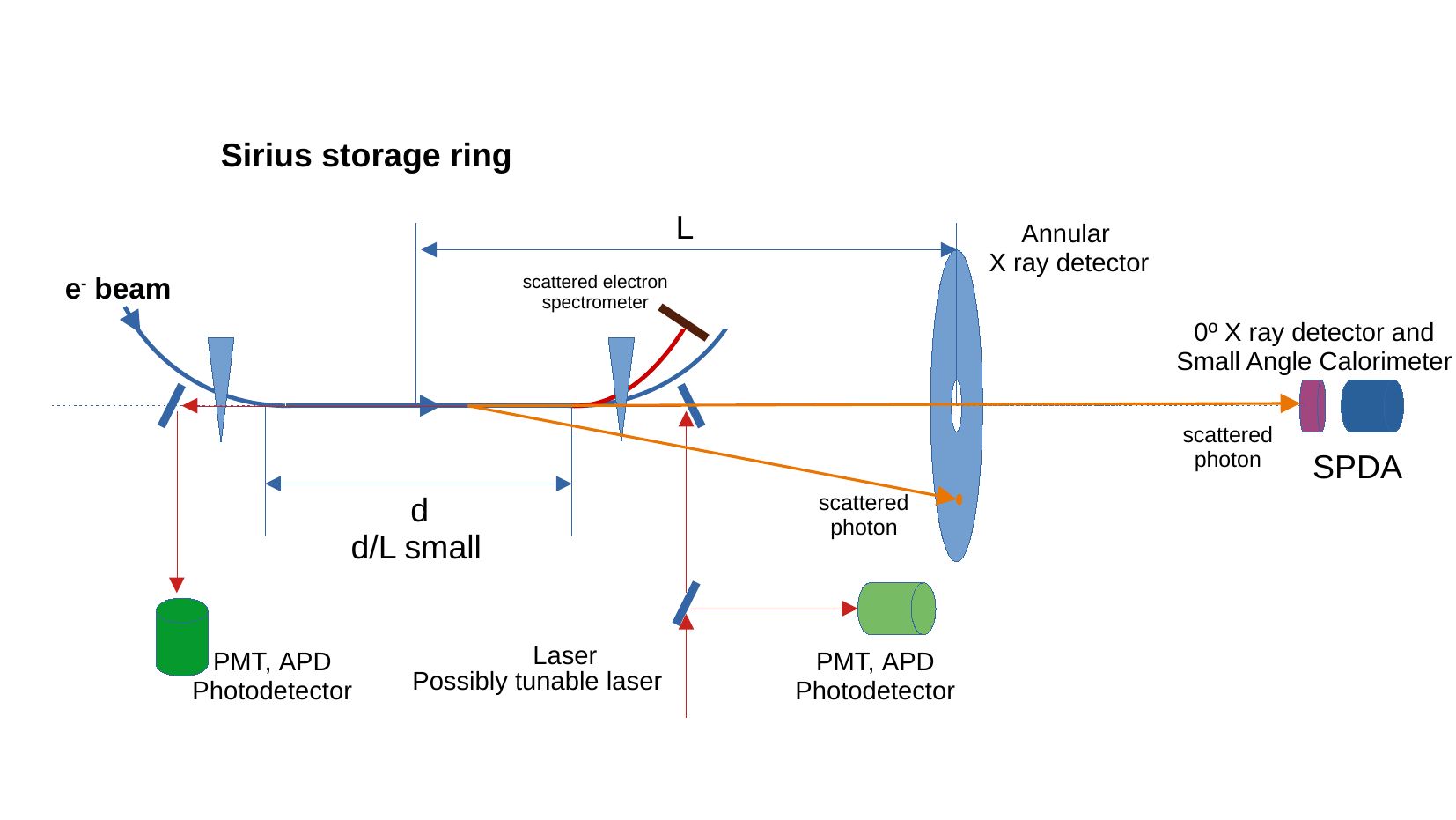}
    \caption{Proposed experimental setup of the experiment illustrating an electron beam (green) with an energy of 3 GeV colliding with a laser of $\mathcal{O}(\mathrm{eV})$ positioned at a certain angle to produce dark photons. The X-ray detectors (and/or calorimeters for high energy photons) detect predominantly background photons.}
    \label{fig:setup2}
\end{figure*}


\noindent The proposed experimental setup, which enables searches for new particle production through DICS, is shown in Figure~\ref{fig:setup2}. The baseline electron-beam, laser, and collision parameters adopted throughout this work are summarized in Table~\ref{tab:beam_laser_parameters}. In particular, we consider a $3~\mathrm{GeV}$ electron beam and $1~\mathrm{eV}$ laser photons in a head-on collision configuration.

For these parameters, the maximum kinematically accessible dark-photon mass is of order
\begin{equation*}
m_{A'}^{\rm max}
\simeq
\frac{2E_{\gamma}E_e}{m_e}
\simeq
\mathcal{O}(10~\mathrm{keV})\,,
\end{equation*}
which restricts the search to the low-mass region.
Since the total available transverse energy is $\leq \mathcal{O}(10~\mathrm{keV})$, the beam electron which served as a scattering target continues to move predominantly in the forward direction together with the rest of the beam particles until the first dipole magnet along the storage ring. Due to its reduced momentum after scattering, it experiences a larger magnetic deflection, enabling its separation from the primary beam envelope. The scattered photons include a high-energy component emitted predominantly forward along the electron beam, and a lower-energy component at larger angles. This motivates the placement of both forward and large-angle photon detectors.

\begin{table}
    \centering

    \renewcommand{\arraystretch}{1.25}
    \begin{tabular}{l l c c}
        \hline
        \textbf{Category} & \textbf{Parameter} & \textbf{Value} & \textbf{Unit} \\
        \hline

        \multicolumn{4}{c}{\textit{Electron beam}} \\
        \hline
        & Beam energy, $E_e$ & $3.0$ & GeV \\
        & Lorentz factor, $\gamma_e$ & $5870.85$ & -- \\
        & Beam current, $I_e$ & $350$ & mA \\
        & Revolution frequency, $f_{\rm rev}$ & $0.5783$ & MHz \\
        & Revolution period, $T_{\rm rev}$ & $1.729$ & $\mu$s \\
        & Number of bunches, $N_b$ & $864$ & -- \\
        & Electrons per bunch, $N_e$ & $4.37\times10^{9}$ & -- \\
        \hline

        \multicolumn{4}{c}{\textit{Electron-beam transverse RMS sizes}} \\
        \hline
        SS-A & $(\sigma_{ex},\sigma_{ey})$ & $(64.9,\;3.4)$ & $\mu$m \\
        SS-B & $(\sigma_{ex},\sigma_{ey})$ & $(19.1,\;2.0)$ & $\mu$m \\
        BC   & $(\sigma_{ex},\sigma_{ey})$ & $(9.5,\;3.5)$ & $\mu$m \\
        \hline

        \multicolumn{4}{c}{\textit{Laser}} \\
        \hline
        & Photon energy, $E_\gamma$ & $1.0$ & eV \\
        & Wavelength, $\lambda_\gamma$ & $\simeq 1240$ & nm \\
        & Pulse repetition rate, $f_\gamma$ & $100$ & MHz \\
        & Pulse energy, $E_{\rm pulse}$ & $200$ & $\mu$J \\
        & Photons per pulse, $N_\gamma$ & $1.25\times10^{15}$ & -- \\
        & Horizontal RMS size, $\sigma_{\gamma x}$ & $\sigma_{ex}$ & -- \\
        & Vertical RMS size, $\sigma_{\gamma y}$ & $\sigma_{ey}$ & -- \\
        \hline

        \multicolumn{4}{c}{\textit{Collision configuration}} \\
        \hline
        & Collision geometry & Head-on & -- \\
        & Collision angle, $\theta_{\rm coll}$ & $180^\circ$ & -- \\
        & Collision frequency, $f_{\rm coll}$ & $100$ & MHz \\
        & Benchmark running time & $365$ & days \\
        \hline
    \end{tabular}

    \caption{
    Baseline electron-beam, laser, and collision parameters adopted in this work.
    The transverse laser dimensions are assumed to be matched to the corresponding electron-beam RMS sizes at the interaction point.
    }
    \label{tab:beam_laser_parameters}
\end{table}

For a head-on collision, the instantaneous luminosity is calculated from the
Gaussian overlap of the electron and laser beams as
\begin{equation}
\mathcal{L}_{\rm inst}
=
\frac{f_{\rm coll}N_eN_\gamma}
{2\pi\Sigma_x\Sigma_y},
\qquad
\Sigma_x=\sqrt{\sigma_{ex}^2+\sigma_{\gamma x}^2},
\qquad
\Sigma_y=\sqrt{\sigma_{ey}^2+\sigma_{\gamma y}^2}.
\label{eq:luminosity}
\end{equation}

With the laser transverse dimensions matched to the electron beam,
$\sigma_{\gamma x}=\sigma_{ex}$ and $\sigma_{\gamma y}=\sigma_{ey}$,
this expression reduces to
\begin{equation}
\mathcal{L}_{\rm inst}
=
\frac{f_{\rm coll}N_eN_\gamma}
{4\pi\sigma_{ex}\sigma_{ey}}.
\end{equation}

The integrated luminosity is obtained as
$\mathcal{L}_{\rm int}=\mathcal{L}_{\rm inst}t_{\rm run}$.
The corresponding luminosities for SS-A, SS-B, and BC are used to calculate
the absolute ICS photon rates shown in Figure~\ref{fig:Fixed_energy_pulse}.

Since the $A'$ remains undetected in the parameter region considered in this work, its production gives rise to a missing-energy signature. We consider two complementary strategies to exploit this signature:

\begin{itemize}

\item A photon-counting strategy based on a single-photon veto. The recoil-electron kinematics are used to determine whether an ICS photon is expected within the fiducial region of the photon detector. The absence of the corresponding photon constitutes the signal signature.

\item A recoil-electron strategy based on the reconstructed electron energy spectrum, with angular information considered as an extension of the analysis.

\end{itemize}

Both strategies therefore exploit the kinematic imbalance generated by replacing the outgoing photon of the corresponding ICS process with an invisible $A'$.

Photon detection is achieved through two complementary systems: first, an annular ring calorimeter, exposed predominantly to photons with energy in the $\mathcal{O}(10-100~\mathrm{keV})$ up to $1~\mathrm{MeV}$ range, to measure the energy spectrum of photons at large angles, and second, single-photon detection array (SPDA) for time-resolved photon counting, coupled to a laser monitoring system.

\par

The SPDA serves as the primary photon-counting system for the single-photon-veto strategy. It can be implemented as a multi-section array, with each section optimized to detect a specific energy range of the recoil photon spectrum, ensuring full coverage from low to high energy photons. By providing time-resolved single-photon counting synchronized with the laser pulses, the SPDA enables precise identification of events in which the photon expected from ICS is not detected. The section devoted to optical photons registration could be realized as a semiconductor device (e.g. array of low-noise high-gain single photon avalanche diodes (SPADs) \cite{5772685}, e.g. silicon photomultipliers (SiPMs) \cite{KLANNER201936}).

When operated under reverse voltage above the breakdown threshold, this semiconductor device converts individual photons into measurable electrical signals through avalanche multiplication. This process exploits the photoelectric effect to amplify faint signals, enabling single-photon detection with nanosecond timing resolution.

Alternatively, multi-anode conventional photomultipliers could also be employed for direct optical photon detection at the expense of a larger material budget and possible absorption of part of the X-ray photons.

The X-ray section could be made of a high-density, high-Z semiconductor material like Cadmium Telluride (CdTe) or Cadmium Zinc Telluride (CZT), attached to a photodetector, or a high-efficiency inorganic scintillator like CsI(Tl) \cite{XrayPlastic:2023}. These materials offer superior absorption efficiency for X-rays compared to plastics, due to their higher atomic number ($Z$) and density \cite{Kumar2021ARO}.

However, plastic scintillators based on bulk polymerization feature high light yield and uniformity and have the advantage of fast emission which could allow better double pulse separation and high rate capabilities.

The higher photon energy section could be made as a conventional crystal electromagnetic calorimeter detecting photons in the $\mathcal{O}(10~\mathrm{MeV}) - 1~\mathrm{GeV}$ range. Inorganic scintillators such as BGO ($\text{Bi}_4\text{Ge}_3\text{O}_{12}$), LSO, LaBr$_3$ and CeBr$_3$ are appropriate for $\mathcal{O}(\mathrm{MeV})$ photon energy due to their high light yield and high $Z$ value, which enhances detection efficiency for Gamma-rays \cite{TakayukiYANAGIDA2018PJA9402B-02}, while for $E_{\gamma}>50~\mathrm{MeV}$ Cherenkov light crystals like PbF$_2$ are the proper choice, allowing single gamma counting with sub-ns time resolution and pulse duration \cite{Rodnyi2020PhysicalPI,Cemmi_2022}.

For the single-photon-veto strategy, the expected ICS photon is determined event by event from the reconstructed recoil-electron kinematics. A signal candidate corresponds to an event in which the associated photon is expected to enter the fiducial photon-detector region but is not observed.
The calorimeter simultaneously records the photon energy distribution, allowing cross-verification of spectral anomalies.

The photons with larger energy will be accompanied by higher energy loss of the corresponding target electron.
Thus the recoil electron will possess lower energy than the nominal Sirius beam energy and will be deflected stronger by the first dipole magnet after the Compton scattering region, allowing its detection with a segmented charged particle detector. Silicon pixel detectors with high resolution, e.g. Timepix3-based arrays \cite{Bertelli:2024jzb,NIEVAART2025170643} or monolith active pixel sensors \cite{949010,Turchetta:712036,4437188} could provide charged particle detection with $\mathcal{O}(\mathrm{ns})$ time resolution and $50~\mu\mathrm{m}$ pixel size. The smaller the pixel size, the better the $e^-$ resolution would be. Event identification can be performed by matching the electron momentum with the photon energy in the annular calorimeter (less probable) or the small angle calorimeter. The reconstructed recoil-electron energy provides the observable used in the second sensitivity strategy, while the reconstructed electron angle can provide additional discrimination when sufficient angular resolution is available.

The proposed setup also features a set of laser control detectors (APD, PMT or other), allowing to precisely follow the laser beam intensity and to measure the flux of the non-interacted laser photons. This monitoring system will operate continuously, providing flux normalization and tracking pulse-to-pulse variations. It therefore provides an important control of the laser intensity and of possible instrumental variations affecting the photon-counting analysis. This multi-detector approach provides robust information which could constrain the dark photon mass $m_{A^{\prime}}$ and coupling parameter $\varepsilon$, with particular sensitivity to feeble interactions. The combined scattered photon detection system exploits the APD's single-photon counting capability, when operated in Geiger mode, and the calorimeter's energy resolution to isolate potential dark photon signatures from instrumental effects. To fully exploit this capability, the experiment must ensure sensitivity to photon energies extending into the near-infrared and sub-eV regime.

The optical photon detection system is designed to provide broad spectral coverage, ensuring sensitivity to potential photon disappearance events. In addition to SPAD/APD-based sensors and SiPM arrays optimized for visible wavelengths, near-infrared photomultiplier (NIR-PMT) modules may be incorporated to extend efficient single-photon detection down to photon energies of approximately $0.7~\mathrm{eV}$. These devices offer lower dark count rates and improved quantum efficiency in the NIR compared to room-temperature semiconductor photodiodes, requiring only moderate cooling (e.g., liquid-nitrogen temperatures). Together with the semiconductor photon counters, they establish the effective low-energy detection threshold of the optical system. Sensitivities below this range require cryogenic detector technologies, whose implementation and performance considerations are discussed in Section~\ref{Lower_Threshold}. This layered approach ensures continuous coverage from optical to sub-eV photon energies and defines the thermal and cryogenic requirements of the overall detection strategy. This extended spectral coverage is essential for identifying potential missing-energy signatures associated with dark photon production.

In particular, the combination of photon-veto information with the reconstructed recoil-electron observables provides complementary probes of the missing-energy signature associated with DICS. This method builds upon established missing-energy techniques used in supersymmetry and dark matter searches \cite{PhysRevD.100.055002,CANEPA2019100033, Abdughani2019}, but adapts them to the unique capabilities of synchrotron light sources. These measurements form the basis of the two sensitivity strategies developed in Section~\ref{Discussion}.

\section{Lower Photon Detection Thresholds}
\label{Lower_Threshold}

To enable the detection of photon disappearance events indicative of dark photon production, the key technological challenge is to achieve high-efficiency photon counting at low energies, particularly in the near-infrared (NIR), mid-infrared (MIR), and extending into the terahertz (THz) spectral range. This requires cryogenic operation and a careful selection of detector architectures optimized for different parts of the energy spectrum.

As discussed in the experimental setup, the optical detection system establishes the effective low-energy threshold of the experiment. In the initial stage, for photon energies down to approximately $0.7~\mathrm{eV}$ (corresponding to wavelengths up to $\approx 1.7~\mu\mathrm{m}$), we propose the use of a near-infrared photomultiplier module. Such devices provide efficient single-photon detection above this threshold with moderate cooling (e.g., liquid-nitrogen temperatures) and complement semiconductor photon counters used at visible wavelengths. This provides a robust and commercially available platform for performing initial photon-counting measurements. Below this energy range, the quantum efficiency of photomultiplier tubes rapidly decreases near the long-wavelength cutoff, making superconducting cryogenic detectors necessary to extend the sensitivity toward lower photon energies.

For the $0.1 - 0.7~\mathrm{eV}$ range, Superconducting Nanowire Single-Photon Detectors (SNSPDs) present a highly promising option. SNSPDs based on materials such as tungsten silicide (WSi) or molybdenum silicide (MoSi) exhibit excellent quantum efficiency in the NIR and MIR range, with detection efficiencies up to 90\% and dark count rates below 1 count per second when cooled to sub-Kelvin temperatures. Their fast response times ($< 100~\mathrm{ps}$) make them well-suited for time-resolved photon counting. Although standard SNSPDs become inefficient below $\sim 0.1~\mathrm{eV}$, ongoing developments in material engineering and antenna-coupled geometries are progressively pushing this limit toward lower energies.

For the $0.01 - 0.1~\mathrm{eV}$ regime, corresponding to far-infrared and THz photons, two alternative technologies are under consideration: Kinetic Inductance Detectors (KIDs) and Superconducting Tunnel Junctions (STJs). KIDs are intrinsically multiplexable, cryogenic detectors that can be engineered for broadband absorption and are particularly effective for THz photon detection. Their typical energy thresholds extend down to a few meV, making them ideal for waveguide- or antenna-coupled setups in a dilution refrigerator. While their time resolution ($\sim \mu\mathrm{s}$) is lower than that of SNSPDs, their scalability and low noise performance make them well-suited for large-area, background-limited measurements.

STJs offer a complementary approach, providing intrinsic energy resolution and single-photon sensitivity down to photon energies of $\sim \mathrm{meV}$. Their operation relies on detecting quasiparticles generated by photon absorption in a superconducting junction, yielding energy-resolved spectra that can help discriminate signal from background. However, they suffer from longer integration times and more complex readout electronics, making them more suitable for targeted energy-tagging applications rather than high-rate detection.

Together, these technologies can be integrated into a tiered detection strategy that covers the sub-eV energy range down to the meV scale. 
In such a configuration, SNSPDs would handle higher-energy (NIR-MIR) photons, while KIDs and STJs would cover the lower-energy THz regime. Note that while the NIR measurements can be performed off axis, for the even lower detection energies waveguides can be efficiently used to inject the photons into the cryogenic setup, which would constitute a larger detector system than the NIR-PMT setup. This multimodal approach would exploit cryogenic quantum-detector technologies to provide broad spectral coverage and enable the identification of photon disappearance events over an extended energy range.

The motivation for extending the photon detection threshold toward the sub-eV
range is directly connected to the angular structure of \textcolor{purple}{ICS}
shown in Fig.~\ref{fig:Fixed_energy_pulse}. In the laboratory frame, the emitted photons are strongly
forward-collimated, with the highest energies concentrated inside the relativistic
cone of opening angle $\sim 1/\gamma$, where $\gamma$ is the Lorentz factor of the incoming electron. Away from this cone, the scattered-photon
energy decreases rapidly, producing a broad population of lower-energy photons
at larger angles. Consequently, increasing the angular acceptance of the detector
alone does not significantly enhance the measurable photon flux unless the
detector remains efficient at low energies.

Lowering the detection threshold therefore enlarges the experimentally accessible photon phase space by allowing soft photons emitted at larger angles to be detected. This extended coverage can provide additional photon-counting information and improve the characterization and monitoring of the ICS photon distribution. However, the sensitivity projections presented in this work do not rely directly on the full low-energy photon population. The single-photon-veto strategy uses a fiducial sample of hard ICS photons for which the corresponding photon is expected to enter the active detector region, whereas the complementary strategy is based on the reconstructed recoil-electron energy spectrum, with angular information included as an extension. The low-energy photon-detection systems discussed in this section should therefore be regarded as an extension of the experimental capabilities of the proposed setup rather than as a requirement for the benchmark sensitivity projections in the following section.\\

\section{\label{Discussion}Discussion}
\subsection{Background}
\label{sec:background}

\begin{figure}[H]
  \centering

  \begin{minipage}{0.88\linewidth}
    \centering
    \includegraphics[width=\linewidth]{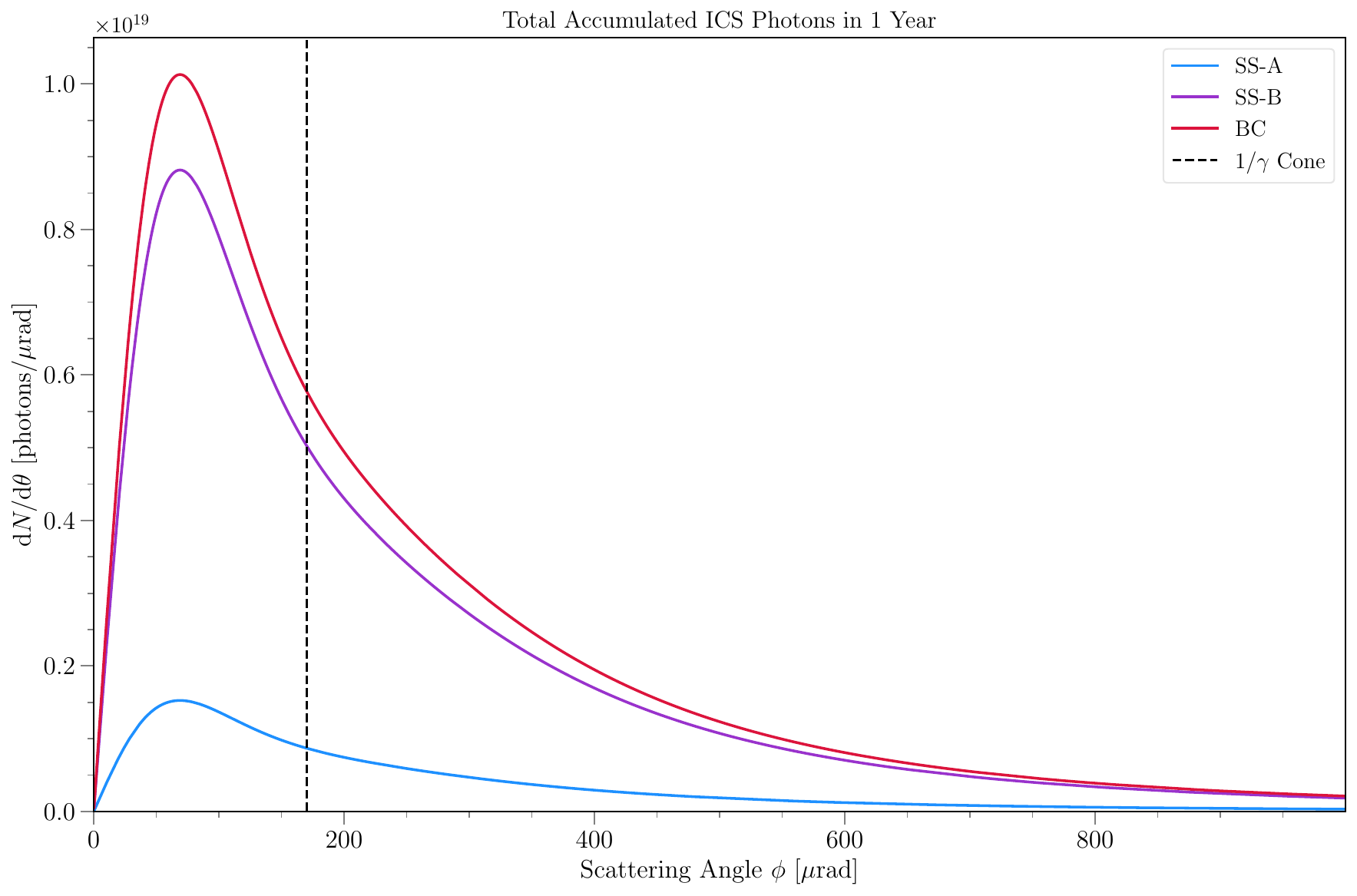}

    \vspace{0.4em}

    {\small (a) Total accumulated photons in 1 year}
  \end{minipage}

  \vspace{1.2em}

  \begin{minipage}{0.88\linewidth}
    \centering
    \includegraphics[width=\linewidth]{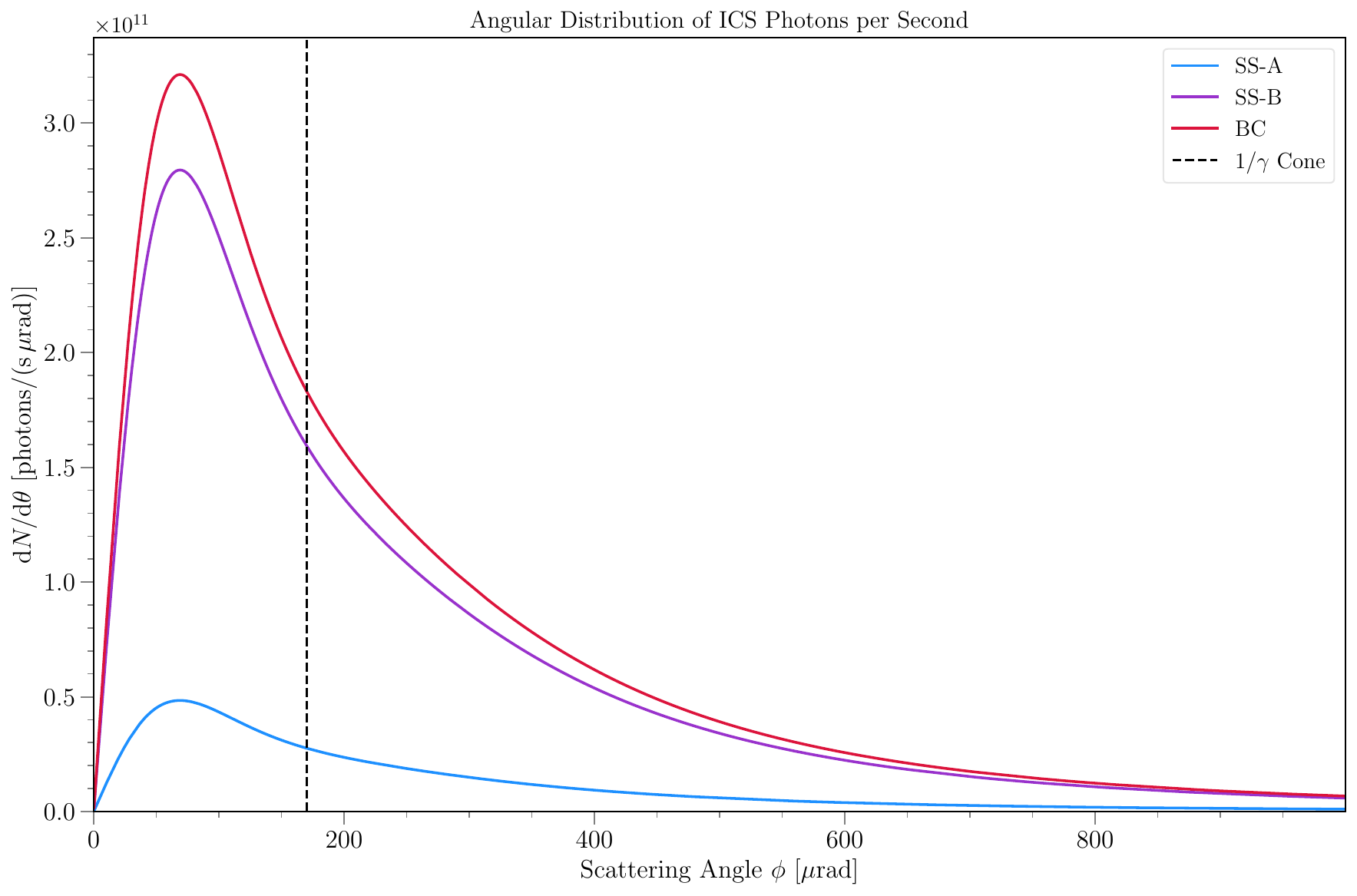}

    \vspace{0.4em}

    {\small (b) Photon rate per second}
  \end{minipage}

  \caption{
  Angular distribution of photons from SM ICS for the three Sirius interaction regions considered in this work: SS-A, SS-B, and BC. The corresponding electron-beam transverse RMS sizes are those listed in Table~\ref{tab:beam_laser_parameters}, with the laser transverse dimensions matched to the electron-beam sizes at each interaction point. Panel (a) shows the total accumulated photons for a data-taking period of 1 year, while panel (b) displays the corresponding photon rate per second. The photon distribution is strongly forward-peaked, with \(1/\gamma \sim 170\,\mu\mathrm{rad}\) providing the characteristic relativistic angular scale. The strong forward collimation of the ICS photons is an important feature for the photon-veto strategy discussed below.} 
  \label{fig:Fixed_energy_pulse}
\end{figure}

The dominant background for both proposed strategies is the SM ICS process, \(e^-\gamma \to e^-\gamma\). This process produces a high flux of photons and recoil electrons with well-defined energy and angular distributions that depend only on the beam and laser parameters. The outgoing photons are strongly collimated in the forward direction, with the majority of the flux confined within the relativistic cone of opening angle \(1/\gamma \sim 170\,\mu\)rad, as shown in Figure~\ref{fig:Fixed_energy_pulse}. The photon distribution extends to energies up to \(\sim 132\,\text{MeV}\)~\cite{Ginzburg:1981vm}, while the recoil electrons carry the complementary energy and remain strongly collimated along the beam direction.

The role of the ICS background differs between the two experimental strategies. In the photon-veto approach (Section~\ref{sec:veto-strategy}), the relevant background arises from ICS events in which the photon expected from the reconstructed recoil-electron kinematics is not detected. The corresponding background rate is therefore controlled by the photon-veto miss probability \(p_{\rm miss}\), together with the event occupancy and the fiducial photon selection. These quantities are defined in Section~\ref{sec:veto-strategy}.

In the recoil-electron approach (Section~\ref{calorimeter_strategy}), the background is the continuous energy (and angular) spectrum of the electrons from ordinary ICS itself. In this case, the signal is not a deficit but a spectral distortion: the DICS process replaces the final-state photon with an invisible dark photon, altering the electron kinematics relative to the pure SM prediction. The background is therefore the predicted QED recoil-electron spectrum, which serves as the baseline against which any deviation is measured. 
For the benchmark sensitivity projection, the ICS spectral shape is taken from the QED prediction, while its global normalization is assigned a representative
1\% Gaussian uncertainty and profiled as a nuisance parameter, as described in Section ~\ref{calorimeter_strategy}. The fixed-normalization case, $\sigma_a=0$, is shown only as
an idealized reference.

The angular distribution shown in Figure~\ref{fig:Fixed_energy_pulse} illustrates the strongly forward nature of ICS at the benchmark beam and laser energies. For the photon-veto strategy, this determines the region in which the accompanying ICS photon is expected to be detected. In the recoil-electron strategy, the corresponding ICS electron spectrum provides the background template against which the DICS-induced distortion is tested. The two approaches therefore exploit complementary manifestations of the same missing-energy signature.

\subsection{Projected Sensitivity}\label{sec:sensitivity}
 
To calculate the projected sensitivity of our proposed setup, we consider two complementary experimental strategies for detecting dark photons produced via DICS: a single-photon veto technique, and a measurement of the recoil electron energy spectrum. Both strategies share the same underlying physics described by the DICS cross section of Eq.~\ref{eq:cross}, which is integrated in the center-of-mass (CM) frame, with the final-state four-momenta subsequently boosted to the laboratory frame for the head-on collision geometry of Figure~\ref{fig:setup2}, but differ in what is measured and in how the exposure is defined. At this stage, both sensitivities are presented as idealized benchmarks. Full detector efficiencies, instrumental backgrounds, and non-Gaussian detector-response effects are left for future detector-level studies.

\subsubsection{The Single-Photon Veto Strategy}
\label{sec:veto-strategy}
 
The first strategy relies on a missing-energy and single-photon-veto
technique, exploiting the photon-counting approach discussed in
Section~\ref{Proposal}. Since the dark photon escapes the detector without depositing energy, the experimental signature is a recoil electron with kinematics compatible with a two-body $e^-\gamma\to e^-A'$ interaction, accompanied by the absence of a Compton-scattered photon in the forward veto.

The irreducible background is ordinary ICS, $e^-\gamma\to e^-\gamma$, for the rare case in which the outgoing photon
evades detection. Both the DICS signal and the ICS background are evaluated with the same validated matrix element (Eq.~\ref{eq:cross} for $m_{A'}\neq0$, and its $m_{A'}\to0$ limit for ordinary Compton scattering) so that the two processes agree exactly in the massless limit.

The expected signal counts are
\begin{equation}
N_{\rm sig}(m_{A'},\varepsilon) \;=\; \varepsilon^2\,\mathcal{L}_{\rm int}\,
\sigma_{\rm sig,fid}(m_{A'},\varepsilon{=}1),
\label{eq:nsig}
\end{equation}
where $\sigma_{\rm sig,fid}$ is the fiducial DICS cross section evaluated at unit kinetic mixing. The background from ICS photons that evade the veto is evaluated at the crossing level. For a Poisson mean $\mu$ of fiducial ICS photons per crossing, the probability that a given crossing contains at least one photon and that all such photons are missed is $e^{-\mu}(e^{\mu p_{\rm miss}} - 1)$. Thus, over the total number of crossings $N_{\rm crossings}=f_{\rm coll}T$, the expected background is
\begin{equation}
B(p_{\rm miss}) = N_{\rm crossings}\, e^{-\mu}\bigl(e^{\mu p_{\rm miss}} - 1\bigr),
\qquad
N_{\rm crossings} = f_{\rm coll} T.
\label{eq:background}
\end{equation}

We estimate the projected $95\%$~C.L. sensitivity using a one-sided Poisson Asimov construction for the median expected upper limit. For a known background $B$, the upper signal count $S_{\rm lim}$ is defined by
\begin{equation}
Q(B+1, B+S_{\rm lim}) = \alpha,\qquad \alpha=0.05,
\label{eq:poisson_asimov}
\end{equation}
where $Q(n,x)=\Gamma(n,x)/\Gamma(n)$ is the regularized upper incomplete gamma function. This construction yields $S_{\rm lim}=-\ln\alpha\simeq 2.9957$ in the background-free limit and approaches the familiar Gaussian convention for large $B$. The corresponding limit on the kinetic mixing follows from
\begin{equation}
\varepsilon_{95}(m_{A'}) \;=\;
\sqrt{\frac{S_{\rm lim}}{\mathcal{L}_{\rm int}\,\sigma_{\rm sig,fid}(m_{A'},\varepsilon{=}1)}}.
\label{eq:eps95}
\end{equation}

Because the matrix element is Lorentz invariant, the cross section is computed as an exact integral over the CM scattering angle and only afterwards the final-state kinematics are boosted to the laboratory frame via the boost appropriate to the head-on collision. The fiducial region entering both $\sigma_{\rm sig,fid}$ and $\sigma_{\rm ICS,fid}$ is defined entirely on these boosted, laboratory-frame quantities:
\begin{itemize}
\item the outgoing boson (photon or $A'$) laboratory energy satisfies
      $E_X \geq 10~\text{MeV}$;
\item its laboratory polar angle satisfies $\theta_X \leq 600~\mu\text{rad}$.
\end{itemize}
For $m_{A'}\to0$ this reproduces the ordinary Compton spectrum shown in
Figure~\ref{fig:Fixed_energy_pulse}, with a fiducial acceptance of $89.2\%$ of the total ICS cross section, $\sigma_{\rm ICS,fid} = 5.67\times10^{11}~\text{pb}$ out of $\sigma_{\rm ICS,total} = 6.36\times10^{11}~\text{pb}$.

\begin{figure}[h!]
    \centering
    \includegraphics[width=1\linewidth]{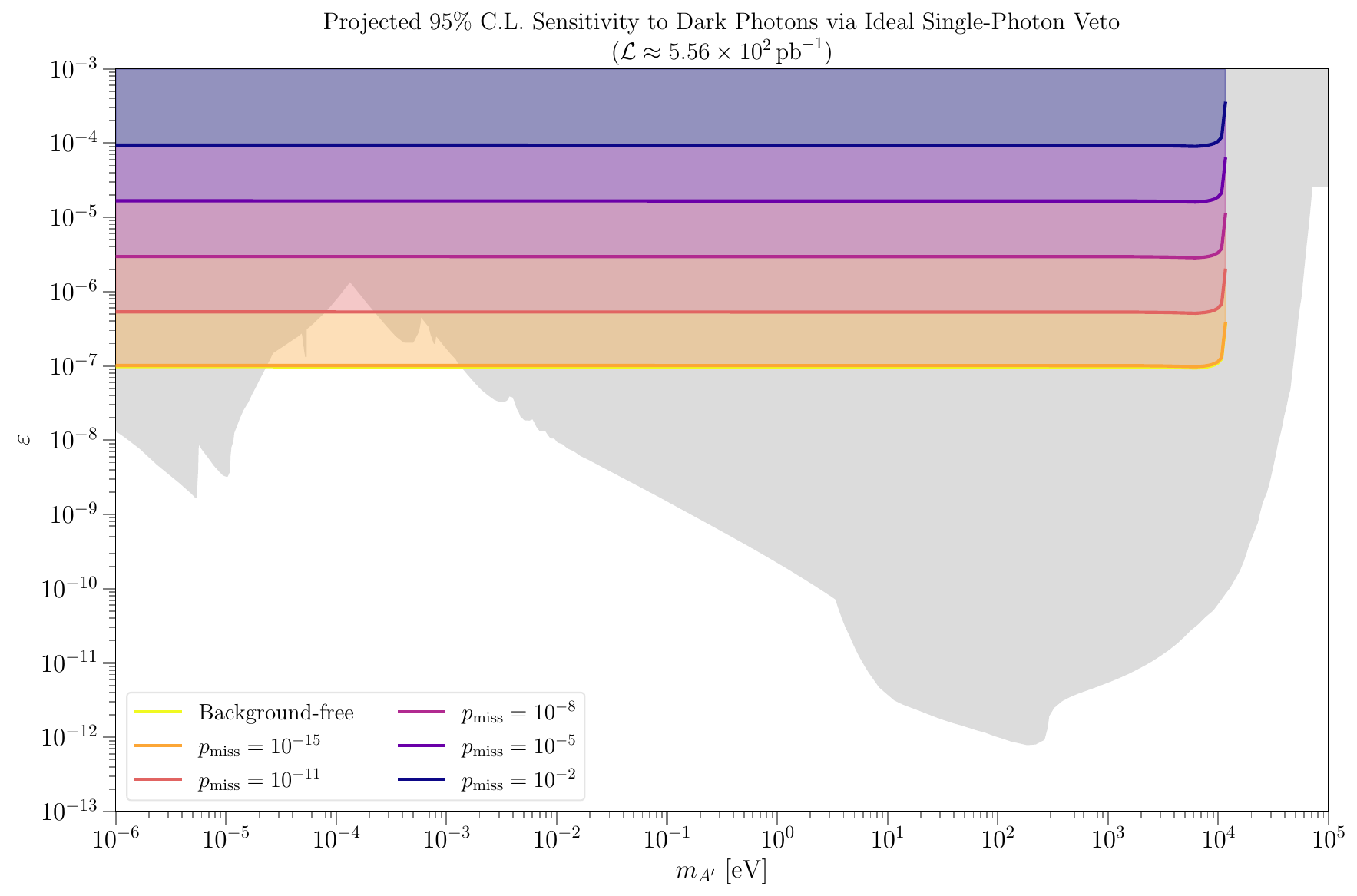}
    \caption{Projected 95\% C.L. sensitivity to dark photons from an ideal single-photon veto analysis, for an integrated luminosity $\mathcal{L} \approx 5.56\times10^{2}\,\mathrm{pb}^{-1}$. Colored curves correspond to different assumed photon-veto miss probabilities $p_{\mathrm{miss}}$, with the black curve showing the background-free benchmark. Gray-shaded regions show existing laboratory constraints.}
    \label{fig:approach1_ideal_veto}
\end{figure}

The very small values of $p_{\rm miss}$ considered here apply only to the fiducial hard-photon sample, rather than to the complete ICS spectrum. The recoil-electron momentum is used on an event-by-event basis to select configurations for which, under the SM Compton hypothesis, the associated photon is expected to satisfy $E_\gamma > 10~{\rm MeV}$ and $\theta_\gamma < 600~\mu{\rm rad}$, and to intercept a fully active region of the forward calorimeter. Events for which the corresponding photon would have an energy below this threshold or propagate outside the fiducial angular region are excluded from the single-photon-veto analysis. For the accepted hard-photon sample, a sufficiently deep high-$Z$ calorimeter can make the intrinsic probability for a photon to traverse the active material without initiating an electromagnetic interaction extremely small.

As an illustrative high-energy estimate, a $40~{\rm cm}$ PbWO$_4$ calorimeter corresponds to approximately $45\,X_0$, where $X_0$ is the radiation length of PbWO$_4$. This regime, in which pair production dominates, gives an asymptotic nonconversion probability of order $10^{-15}$. The actual photon interaction probability over the full $10$--$132~{\rm MeV}$ fiducial energy range must be evaluated using the energy-dependent photon attenuation cross section. Moreover, the experimentally relevant veto miss probability $p_{\rm miss}$ is not determined by the intrinsic interaction probability alone, but also includes geometrical acceptance, inactive regions, readout failures, reconstruction inefficiencies, and related detector effects. We therefore treat $p_{\rm miss}$ as an experimental performance parameter and present projected sensitivities for a range of assumed values, rather than assigning a specific full-detector inefficiency.

The nominal luminosity obtained from the beam and laser parameters summarized in Table~\ref{tab:beam_laser_parameters} would correspond to approximately $\mathcal{O}(10^5$--$10^6)$ fiducial ICS photons per crossing, depending on the Sirius interaction region. Such occupancies are far beyond the regime in which a single-photon veto can resolve individual crossings. We therefore fix instead a target mean occupancy $\mu$ of fiducial ICS photons per crossing and derive the corresponding exposure from it. This choice of operating point corresponds to the laser intensity at which the veto can reliably discriminate between crossings with zero photons and those with one or more photons, rather than being a prediction of the nominal Sirius luminosity. Setting $\mu = 0.1$ keeps the probability of more than one fiducial ICS interaction per crossing below $0.5\%$, and gives
\begin{equation}
\mathcal{L}_{\rm int} \;=\; \frac{\mu\, f_{\rm coll}\, T}{\sigma_{\rm ICS,fid}},
\label{eq:exposure}
\end{equation}
with $f_{\rm coll} = 100$~MHz and $T$ one calendar year at $100\%$ duty cycle. This yields $\mathcal{L}_{\rm int} \approx 556~\text{pb}^{-1}$ and $N_{\rm ICS,fid} = \mu f_{\rm coll} T \approx 3.15\times10^{14}$ expected fiducial ICS photons over the full exposure.

Figure~\ref{fig:approach1_ideal_veto} shows $\varepsilon_{95}(m_{A'})$ for $p_{\rm miss} \in \{10^{-15},\,10^{-11},\,10^{-8},\,10^{-5},\,10^{-2}\}$, together with the background-free benchmark $p_{\rm miss}=0$. Because $\sigma_{\rm sig,fid}$ varies only weakly with $m_{A'}$ below the kinematic endpoint, each curve is nearly flat until the sharp upturn near $m_{A'}\simeq m_{A',\max}\simeq11.6$~keV, where the two-body phase space closes. The DICS cross section approaches the ICS one as $m_{A'}\to0$, as illustrated in Figure~\ref{fig:cs_mDP}. The vertical separation between the curves is set entirely by $p_{\rm miss}$. For $p_{\rm miss}\lesssim \text{a few}\times10^{-15}$, the expected background remains of order one event or less over the full exposure and the analysis approaches the background-free regime, with $\varepsilon_{95}\approx9.7\times10^{-8}$ at low mass. For the larger values of $p_{\rm miss}$ shown in Figure~\ref{fig:approach1_ideal_veto}, the background from missed ICS photons increasingly dominates and the reach weakens accordingly. In this regime $\varepsilon_{95}\propto B^{1/4}$ at fixed luminosity, which implies $\varepsilon_{95}\propto\mathcal{L}_{\rm int}^{-1/4}$, in contrast to the background-free scaling $\varepsilon_{95}\propto\mathcal{L}_{\rm int}^{-1/2}$. The probability $p_{\rm miss}$ that a real, in-acceptance Compton photon fails to register in the veto is therefore the single most consequential experimental parameter for this strategy. The curves shown should be read as scenario bands spanning this quantity, not as statistical uncertainty bands. The gray shaded region reproduces existing laboratory constraints on $(m_{A'},\varepsilon)$, as in Figure~\ref{fig:Upper_Limits}.

This result is an idealized benchmark: it assumes perfect recoil-electron reconstruction, a constant and perfectly known $p_{\rm miss}$, and no background beyond missed ICS photons. In practice $p_{\rm miss}$ will depend on the photon energy, incidence angle, and position on the veto. The background estimate already accounts for pile-up through the per-crossing factor in Eq.~\ref{eq:background}. The main remaining idealization is the assumption of unit signal efficiency; accidental losses of genuine dark-photon signal events due to overlapping ICS photons in the same crossing are not modeled and are deferred to future detector-level studies. We compare this strategy to the recoil-electron-spectrum approach, which adopts a lower target ICS occupancy and a correspondingly different analysis exposure, in Section~\ref{calorimeter_strategy}.

\subsubsection{The Recoil-Electron Energy Spectrum Measurement}
\label{calorimeter_strategy}

The second strategy relies on a precise measurement of the recoil-electron energy
spectrum. Unlike the photon-veto approach, which searches for
events in which the photon expected from the recoil-electron kinematics is not
detected, this strategy uses ordinary ICS as a well-known
continuous background. If dark photons are produced, the DICS process replaces
the final-state photon with an invisible $A'$, altering the energy (and angular)
distribution of the recoil electron. The signal thus manifests as a spectral
distortion of the electron relative to the pure QED prediction.

The expected background and signal templates are built from the exact matrix
element described in the Single-Photon Veto Strategy. The background template
$B_i$ corresponds to ordinary ICS, while the signal template $\hat S_i$
corresponds to DICS evaluated at unit coupling, $\varepsilon=1$.
To account for the uncertainty in the overall normalization
of the ICS background, we introduce a fractional nuisance parameter $a$. The
expected number of events in bin $i$ is therefore
\begin{equation}
\lambda_i(\varepsilon,a)
=
(1+a)B_i+\varepsilon^2\hat S_i,
\label{eq:lambda_spectrum}
\end{equation}
where $a=0$ corresponds to the nominal ICS normalization,
while the additional term $\varepsilon^2\hat S_i$ accounts for dark-photon production.
As in the photon-veto analysis, the cross sections are
integrated in the CM frame and the final-state kinematics are subsequently
boosted to the laboratory frame.

\begin{figure}
    \centering
    \includegraphics[width=1\linewidth]{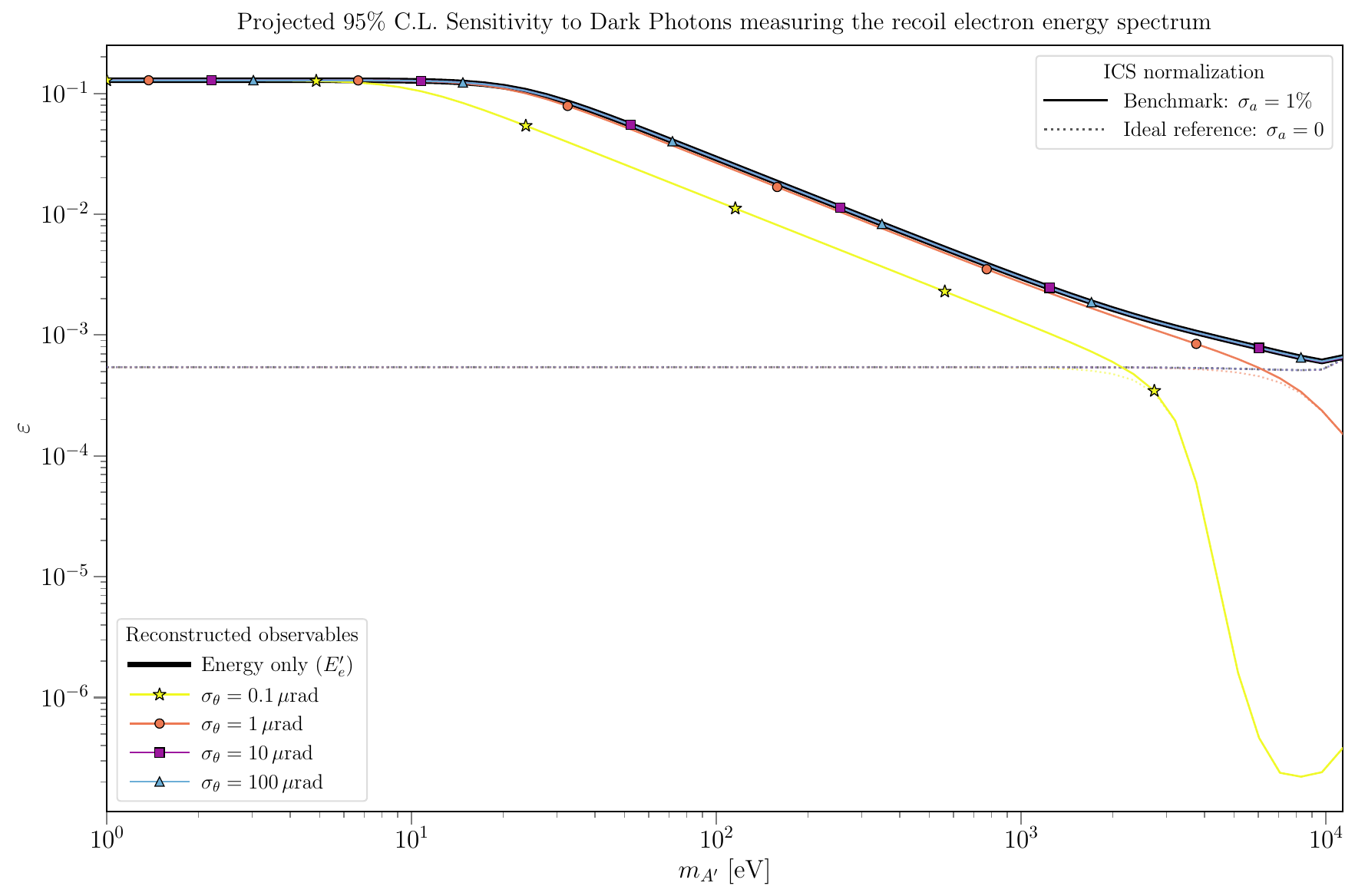}
    \caption{
    Projected 95\% C.L. sensitivity to dark photons from the recoil-electron spectrum analysis, for $\mathcal{L}_{\rm int}\simeq 49.6\,\mathrm{pb}^{-1}$ and $\mu_{\rm ICS}=0.01$ per crossing. Solid curves show the benchmark results with a representative $1\%$ global ICS-normalization uncertainty, while dotted curves show the idealized fixed-normalization case, $\sigma_a=0$. The black curves correspond to the energy-only analysis with $\sigma_E=1\,\mathrm{MeV}$, and the colored curves to the $(E'_e,\theta'_e)$ analysis for $\sigma_\theta=0.1$, $1$, $10$, and $100\,\mu\mathrm{rad}$.
    }
    \label{fig:approach2_Etheta}
\end{figure}

For the energy-only analysis, we use the marginal template in the reconstructed
electron energy $E'_e$. To quantify the potential gain from angular information,
we introduce a parametric extension to two dimensions, $(E'_e,\theta'_e)$.
The two-dimensional template is
\begin{equation}
T_{ij}^{E\theta}
=
\mathcal L_{\rm int}
\sum_n w_n \, P_{in}^E \, P_{jn}^{\theta},
\label{eq:template2d}
\end{equation}
where $P_{in}^E$ is a Gaussian energy response with constant resolution
$\sigma_E=1$~MeV, and $P_{jn}^{\theta}$ is a truncated Gaussian angular
response renormalized for $\theta\ge0$. The angular resolution $\sigma_\theta$
is treated as an external parameter; we scan values from $0.1$ to
$100\,\mu$rad to map the possible improvement.
The same energy resolution and exposure are used for all
curves, so that the comparison isolates the additional information provided
by the reconstructed electron angle. The projection of the two-dimensional
template onto the energy axis reproduces the corresponding one-dimensional
energy template.

As in the photon-veto strategy, the exposure is defined
through an occupancy requirement rather than by using the full nominal
luminosity of the accelerator. For the recoil-electron analysis we adopt a
target mean occupancy of
\begin{equation}
\mu_{\rm ICS}=0.01
\end{equation}
ICS interactions per crossing in order to suppress pile-up
in the electron reconstruction. Using the total ICS cross section
$\sigma_{\rm ICS,total}=6.36\times10^{11}$~pb, the corresponding integrated
luminosity is
\begin{equation}
\mathcal L_{\rm int}
=
\frac{\mu_{\rm ICS}\,f_{\rm coll}\,T}
{\sigma_{\rm ICS,total}}
\simeq
49.6~\mathrm{pb}^{-1}.
\label{eq:exposure_electron}
\end{equation}
This operating point corresponds to reducing the laser
intensity relative to the nominal configuration so that the recoil-electron
spectrum can be reconstructed with sufficiently low event occupancy.

We adopt a 1\% Gaussian uncertainty on the global ICS normalization as a representative systematic benchmark. This value is not derived from a detailed detector-performance, machine-stability, or beam-overlap study and should therefore not be interpreted as an expected experimental precision. The very large ICS control sample makes the purely statistical uncertainty on the overall normalization negligible. In practice, the achievable normalization
precision will instead be determined by systematic effects such as electron- and laser-beam monitoring, interaction-point overlap stability, detector acceptance and reconstruction efficiency, and their associated calibrations.
The ICS normalization can also be constrained in situ using the control sample itself. Consequently, the actual experimental precision could be better than the 1\% benchmark adopted here.

The projected sensitivity is obtained from a binned Poisson likelihood using
the background-only Asimov data set.
For the benchmark results shown in Fig.~\ref{fig:approach2_Etheta}, we take
$\sigma_a=0.01$ and treat the corresponding global ICS normalization
uncertainty as a Gaussian-constrained nuisance parameter.
For a given value of $\varepsilon$, the nuisance parameter is profiled when
evaluating the test statistic. The Asimov test statistic is therefore
\begin{equation}
q_A(\varepsilon)
=
\min_a
\left\{
2\sum_i
\left[
\lambda_i(\varepsilon,a)-B_i
+
B_i\ln\left(
\frac{B_i}{\lambda_i(\varepsilon,a)}
\right)
\right]
+
\left(\frac{a}{\sigma_a}\right)^2
\right\},
\label{eq:q_poisson2}
\end{equation}
with $\sigma_a=0.01$. The sum extends over the reconstructed-energy bins in
the one-dimensional analysis and over the reconstructed energy-angle bins in
the two-dimensional analysis. The final term represents the Gaussian
constraint on the global ICS normalization. The projected $95\%$ C.L. upper
limit is obtained from
\begin{equation}
q_A(\varepsilon_{95})=2.71.
\end{equation}

The impact of this normalization uncertainty depends strongly on the
dark-photon mass. In the quasi-massless limit, the normalized DICS spectrum
becomes nearly degenerate with the ordinary ICS spectrum, so that
$\hat S_i$ is approximately proportional to $B_i$. A dark-photon contribution
can then be partially absorbed by a shift in the nuisance parameter $a$, and
the sensitivity of the spectral analysis becomes controlled by the assumed
normalization uncertainty. At larger $m_{A'}$, the DICS spectrum develops
mass-dependent distortions and kinematic endpoints that cannot be reproduced
by a global rescaling of the ICS background. The normalization degeneracy is
therefore progressively broken and the sensitivity becomes increasingly
driven by spectral-shape information.

To illustrate this effect explicitly, Fig.~\ref{fig:approach2_Etheta} also
shows the idealized fixed-normalization case, $\sigma_a=0$, as dotted curves,
while the solid curves correspond to the representative $\sigma_a=1\%$
benchmark. The separation between the corresponding solid and dotted curves
therefore quantifies the impact of the global ICS normalization uncertainty.
As $m_{A'}$ increases and mass-dependent distortions become more pronounced,
this difference is reduced and the sensitivity becomes increasingly
shape-driven.

Figure~\ref{fig:approach2_Etheta} shows the resulting limits. The black
energy-only curves provide the baseline against which the angular extensions
are compared, while the colored curves show the parametric gain obtained by
including the reconstructed recoil-electron angle for different assumed
values of $\sigma_\theta$.
For angular resolutions much larger than the characteristic
recoil-electron angular scale, the two-dimensional result approaches the
energy-only sensitivity. Improved angular resolution provides additional
discrimination between the ICS and DICS kinematic distributions, with the
largest potential gains obtained for $\sigma_\theta\lesssim1\,\mu$rad.
The $\sigma_\theta=0.1\,\mu$rad case represents an especially optimistic
benchmark and, in regions where the relevant bins contain very small expected
event counts, the asymptotic likelihood approximation should be interpreted
with caution. The angular curves should therefore be understood as parametric
benchmarks illustrating the potential benefit of angular information rather
than as detector-performance projections.

Unlike a purely truth-level calculation, the present
benchmark includes Gaussian smearing of the reconstructed electron energy and,
for the two-dimensional extension, of its polar angle.
It also includes the representative 1\% global ICS
normalization uncertainty described above.
Unit electron-reconstruction efficiency is assumed, and thresholds,
non-Gaussian response tails, energy-angle correlations, acceptance losses,
and machine-induced backgrounds are not included. The calculation therefore
provides a controlled framework for quantifying the potential value of angular
information, which can be refined once a realistic detector response is
available.

Although our reasoning has been based on the coupling of dark photons with
the electromagnetic current, the experimental strategy
could also be adapted to search for axion-like particles
(ALPs) coupled to electrons, as well as other feebly interacting light scalars.
Naturally, the production cross-section of those particles is different. Thus,
an entirely new analysis would have to be performed. We leave this for future
work. Nevertheless, it is important to stress that our work can serve as a
platform to probe light particles with masses in the sub-eV range and below
using accelerators in an orthogonal way to existing proposals
\cite{Sikivie:1993jm,Homma:2014rja,Hasebe:2015jxa,Homma:2022ktv,
Chao:2024owf,Iwazaki:2025ole,Bartnick:2025lbv,Alves:2026sgo}.

\section{Conclusions}
\label{conclusions}

In this work, we have proposed a novel experimental setup for the production and
indirect detection of dark photons via DICS. For concreteness,
we adopted beam parameters representative of the Sirius
accelerator and considered a detector configuration specifically designed for
this search. By exploiting the high-intensity $3~\mathrm{GeV}$ electron beam
and combining it with a $1~\mathrm{eV}$ monochromatic laser, we explored the
production channel $e^-\gamma \rightarrow e^-A'$, where $A'$ denotes an
invisible dark photon. The associated cross section, given in
Eq.~\ref{eq:cross}, exhibits a strong dependence on the dark photon mass
$m_{A'}$ and the collision angle $\theta$, with the largest production cross section obtained for
head-on collisions ($\theta = 180^\circ$), although configurations with $\theta \geq 135^\circ$ remain feasible.

We have computed the cross section for different collision angles and laser
wavelengths, as shown in Figure~\ref{fig:cs_mDP} and
Figure~\ref{fig:angles}, and adopted the head-on configuration as the benchmark
operating point. For a $3~\mathrm{GeV}$ electron beam and
$1~\mathrm{eV}$ laser photons, the maximum kinematically accessible dark-photon
mass is approximately $11.6~\mathrm{keV}$. The variation with laser wavelength
indicates that a $1~\mathrm{eV}$ laser, corresponding to
$\lambda\simeq1240~\mathrm{nm}$, lies within the optimal range for probing the
targeted mass window.

The projected sensitivity was evaluated using two complementary experimental
strategies with exposures determined by their respective occupancy requirements
rather than by the full nominal luminosity of the accelerator. In the
single-photon-veto strategy, a target occupancy $\mu=0.1$ of fiducial ICS photons
per crossing corresponds to an integrated luminosity
$\mathcal L_{\rm int}\simeq5.56\times10^{2}~\mathrm{pb}^{-1}$. For the
recoil-electron-spectrum strategy, a lower occupancy
$\mu_{\rm ICS}=0.01$ gives
$\mathcal L_{\rm int}\simeq4.96\times10^{1}~\mathrm{pb}^{-1}$.

A particularly relevant result of this study is that the single-photon-veto
strategy probes a region of the $(m_{A'},\varepsilon)$ parameter space that is
not covered by the existing laboratory constraints shown in this work. In
particular, for finite dark-photon masses approximately in the range
$10^{-6}~\mathrm{eV}\lesssim m_{A'}\lesssim10^{-4}~\mathrm{eV}$, the projected
sensitivity reaches kinetic mixings in the range
$\varepsilon\sim10^{-7}$--$10^{-6}$.

In this low-mass region, the DICS kinematic distributions have already
approached their quasi-massless limit and become essentially indistinguishable
in shape from those of ordinary ICS. The sensitivity therefore does not arise
from kinematic discrimination. Instead, it originates from the exclusive
missing-photon signature: the reconstructed recoil-electron kinematics are used
event by event to select configurations for which the ordinary ICS hypothesis
predicts a hard photon inside the fiducial calorimeter acceptance, whereas in a
DICS event the $A'$ escapes detection. The experiment can therefore remain
sensitive to invisible dark-photon production while having essentially no
sensitivity to the precise value of $m_{A'}$ in this regime. The approximately
flat sensitivity at low mass should thus be interpreted as the quasi-massless
limit for finite $m_{A'}$, rather than as a kinematic mass measurement or as a
statement about the exactly massless theory.

The recoil-electron energy-spectrum measurement provides a complementary probe.
For this analysis, we adopt a representative 1\% uncertainty on the global ICS
normalization, implemented as a Gaussian-constrained nuisance parameter and
profiled in the likelihood. This value is used as a systematic benchmark and
should not be interpreted as a projected experimental precision. The very
large ICS control sample makes the purely statistical uncertainty on the
overall normalization negligible; the achievable normalization precision will
instead be determined by systematic effects such as electron- and laser-beam
monitoring, interaction-point overlap stability, detector acceptance,
reconstruction efficiency, and calibration.

At sufficiently small $m_{A'}$, where the normalized ICS and DICS kinematic distributions become nearly degenerate, a dark-photon contribution can be largely absorbed into a shift of the overall ICS normalization. Consequently, the low-mass reach of the recoil-electron spectral analysis is controlled by the assumed normalization uncertainty. At larger masses, finite-$m_{A'}$
effects generate spectral distortions and kinematic features that cannot be reproduced by a global normalization shift, thereby breaking this degeneracy.
In this regime, sufficiently precise reconstruction of the recoil-electron angle can provide additional discrimination and improve the sensitivity.

Given that dark photons in the mass range considered here are invisible in the
detector, their detection relies on indirect techniques. We have proposed two complementary strategies exploiting the resulting
missing-energy signature. The first is based on a
single-photon veto: the reconstructed recoil-electron
kinematics are used to determine whether an ordinary ICS photon should enter the
fiducial photon-detector region, and the absence of the expected photon defines
the signal candidate. The second strategy uses the reconstructed recoil-electron energy spectrum, in which
DICS produces a distortion relative to the ordinary ICS prediction.
The recoil-electron angle was additionally investigated as a
parametric extension of this spectral analysis.

The two strategies are limited by different experimental considerations. In the single-photon-veto analysis, the achievable reach depends critically on the effective probability for a hard fiducial ICS photon to escape the veto. In the recoil-electron spectral analysis, the enormous ICS control sample strongly
suppresses statistical limitations, while the ultimate sensitivity depends on the control of the ICS normalization and on the reconstruction of the mass-dependent spectral shape. Detector efficiencies, event occupancy, beam and laser stability, acceptance, and instrumental systematics will therefore ultimately determine the experimentally achievable sensitivity.

The experimental setup could also be adapted to search for
axion-like particles coupled to electrons, as well as other feebly interacting
light states. In such cases, processes of the form
$e^-+\gamma\rightarrow e^-+a$ could generate analogous missing-energy
signatures. However, their production cross sections and
kinematic distributions differ from those of dark photons, and dedicated
sensitivity analyses would therefore be required. We leave such studies for
future work.

Overall, the proposed experiment presents a
complementary accelerator-based approach to existing
searches for light weakly coupled particles. For dark
photons, the benchmark Sirius configuration probes masses up to the
$\mathcal O(10~\mathrm{keV})$ scale.
In particular, the single-photon-veto strategy provides sensitivity to a previously unexplored region around $10^{-6}~\mathrm{eV}\lesssim m_{A'}\lesssim10^{-4}~\mathrm{eV}$ and
$\varepsilon\sim10^{-7}$--$10^{-6}$, depending on the achievable photon-veto performance. The recoil-electron spectral analysis provides a complementary probe at finite masses, with its sensitivity governed by the mass-dependent spectral distortions, the detector resolution, and the achievable control of
the ICS normalization.

The combination of high-intensity electron beams, laser photons,
photon-veto information, and reconstructed recoil-electron observables offers
a new route for exploring weakly constrained regions of dark-sector parameter
space.
\newpage
\appendix

\section{Recoil-Electron Energy Distribution in DICS}
\label{Distribution_electron}

\begin{figure}[H]
    \centering
    \includegraphics[width=0.9\linewidth]{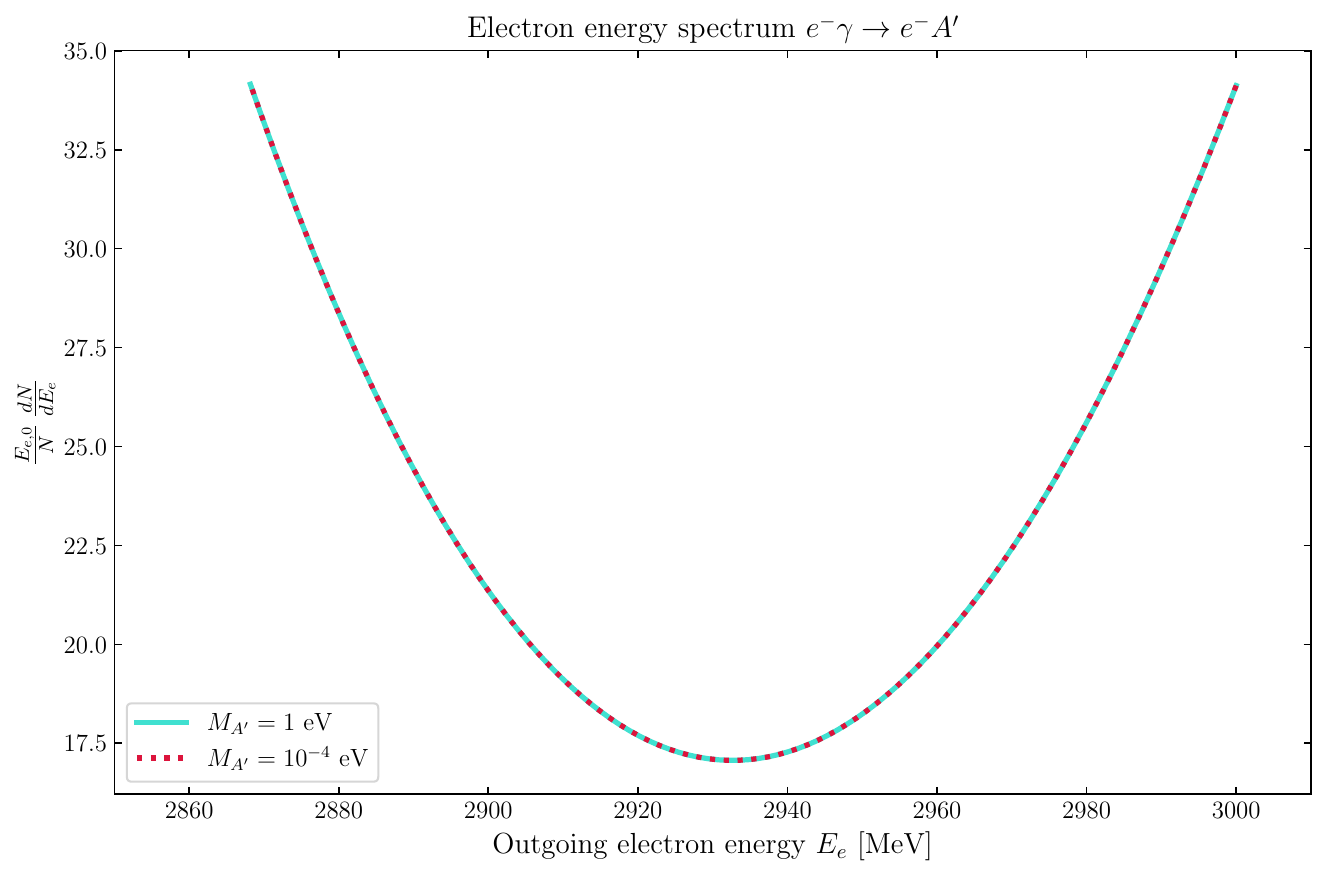}
    \caption{
        Normalized energy spectrum of the outgoing electron in the process
        $e^- \gamma \rightarrow e^- A^\prime$ for
        $m_{A'} = 1~\mathrm{eV}$ and
        $m_{A'} = 10^{-4}~\mathrm{eV}$,
        with kinetic mixing $\varepsilon = 10^{-7}$ and incident electron
        energy $E_{e,0}=3~\mathrm{GeV}$.
        The two spectra are practically indistinguishable, illustrating the
        quasi-massless regime in which the normalized DICS kinematics approach
        those of ordinary ICS.
    }
    \label{fig:outgoing-electron-spectrum}
\end{figure}

The energy spectrum of the outgoing electron in the process
$e^- \gamma \rightarrow e^- A^\prime$
is shown in Fig.~\ref{fig:outgoing-electron-spectrum}.
A significant fraction of the events corresponds to recoil electrons with an
energy loss of $\mathcal{O}(100~\mathrm{MeV})$ relative to the incident
$3~\mathrm{GeV}$ beam energy. Under the ordinary ICS hypothesis, such an energy
loss is accompanied by a hard final-state photon, which should be detected with
very high efficiency in the forward calorimeter. In a DICS event, instead, the
corresponding energy is carried by the invisible $A'$, providing the basis for
the missing-photon signature used in the single-photon-veto strategy.

In particular, for photons in the hard part of the ICS spectrum, the intrinsic
probability of escaping a sufficiently deep high-$Z$ electromagnetic calorimeter
without interacting can be extremely small. As an illustrative high-energy
estimate, a $40~{\rm cm}$ PbWO$_4$ calorimeter corresponds to approximately
$45\,X_0$, where $X_0$ is the radiation length of PbWO$_4$. In the regime where
pair production dominates, the photon interaction length approaches
$\lambda_\gamma \simeq 9X_0/7$, such that the probability of traversing the full
calorimeter depth without conversion is approximately
\begin{equation}
P_{\rm no\,int}
\simeq
\exp\left[-\frac{7d}{9X_0}\right],
\end{equation}
giving $P_{\rm no\,int}\sim10^{-15}$ for $d\simeq45X_0$. This estimate concerns
only the intrinsic noninteraction probability and should not be identified with
the total experimental photon-veto inefficiency. In particular, close to the
lower analysis threshold of $E_\gamma=10~{\rm MeV}$, the asymptotic
pair-production approximation is not sufficient and the energy-dependent total
photon attenuation cross section must be used.

The experimentally relevant miss probability $p_{\rm miss}$ additionally
incorporates inactive material, detector gaps, readout and reconstruction
failures, timing requirements, and other instrumental effects. Consequently,
the achievable value of $p_{\rm miss}$ must ultimately be established
experimentally using the very large simultaneously recorded ICS control sample.

For sufficiently small $m_{A'}$, the normalized recoil-electron spectrum
approaches the ordinary ICS result and becomes essentially independent of the
dark-photon mass. This is the quasi-massless regime discussed in
Section~\ref{Discussion}: although the experiment can remain sensitive to
invisible production through the missing-photon signature, the recoil-electron
kinematics alone contain essentially no information on the precise value of
$m_{A'}$.

\section{Flux of Dark Photons as a Function of the Scattering Angle
\texorpdfstring{$\phi$}{phi}}
\label{Flux_DP}

\begin{figure}[H]
  \centering

  \begin{minipage}{0.86\linewidth}
    \centering
    \includegraphics[width=\linewidth]{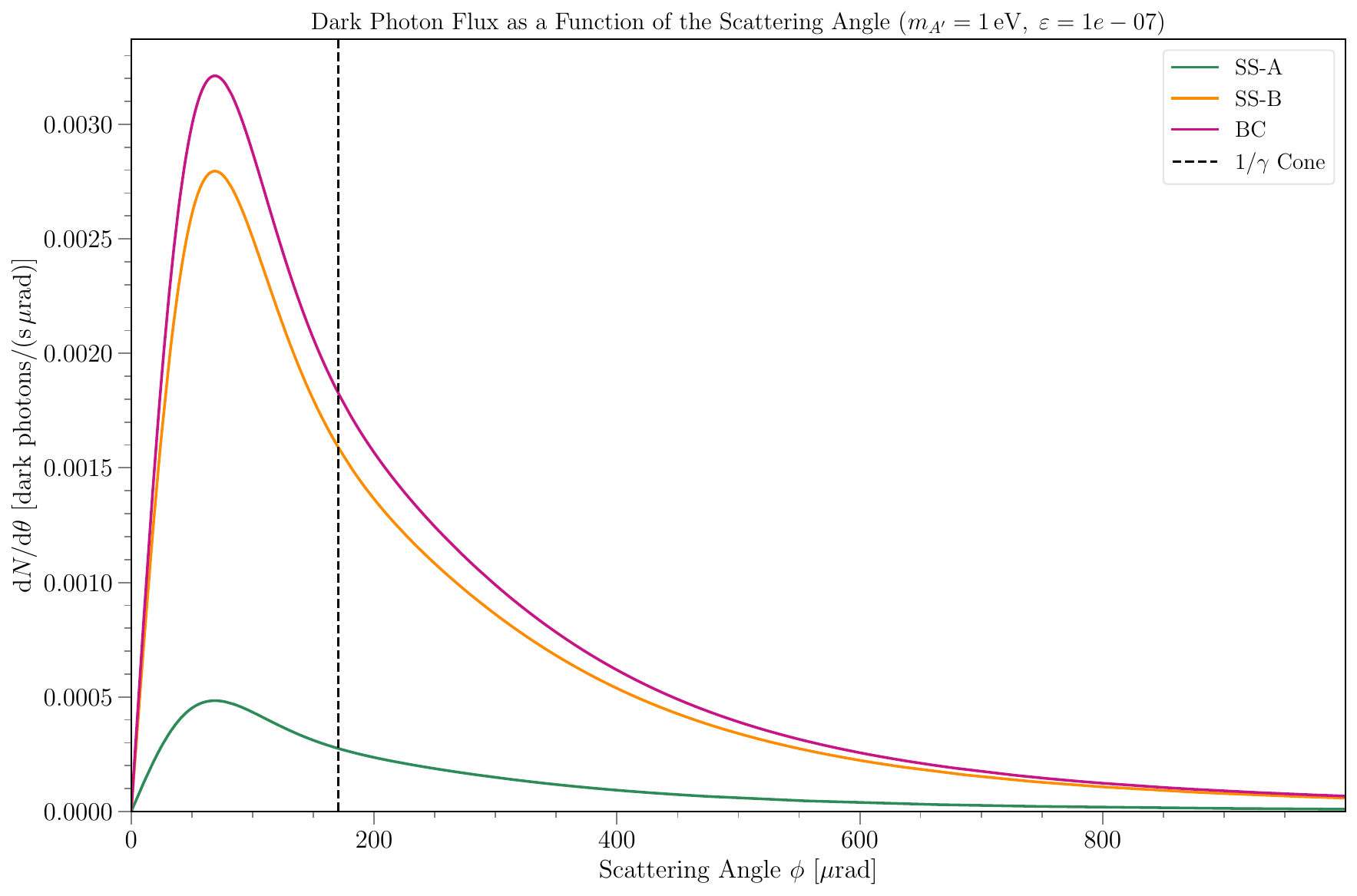}

    \vspace{0.4em}

    {\small (a) Dark photon rate per second}
  \end{minipage}

  \vspace{1.2em}

  \begin{minipage}{0.86\linewidth}
    \centering
    \includegraphics[width=\linewidth]{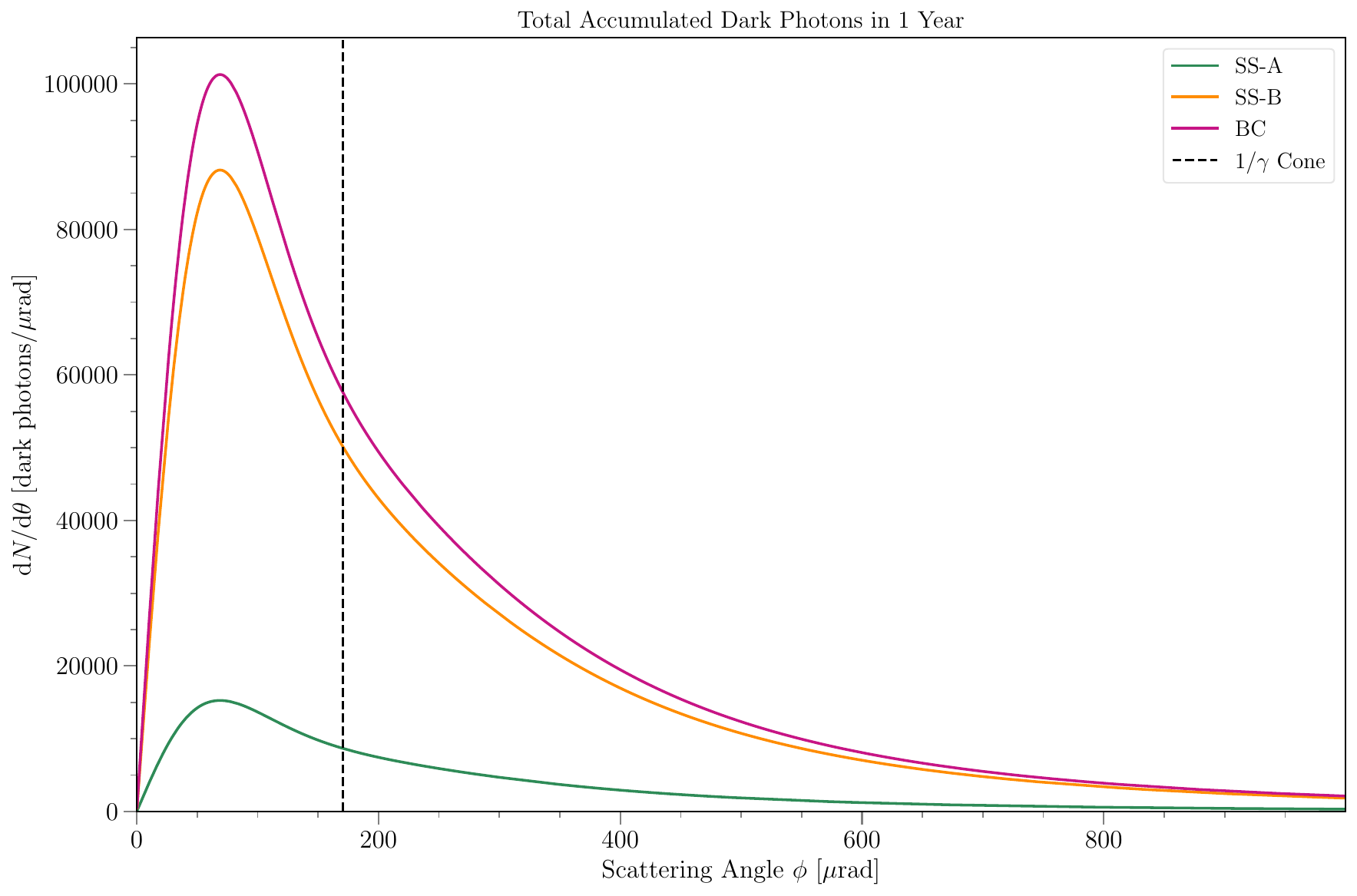}

    \vspace{0.4em}

    {\small (b) Total accumulated dark photons in 1 year}
  \end{minipage}

  \caption{
  Angular distribution of dark photons produced through DICS
  for \(m_{A^\prime}=1~\mathrm{eV}\) and
  \(\varepsilon=10^{-7}\).
  Panel (a) shows the differential dark-photon production rate per second,
  while panel (b) shows the total number of dark photons produced after
  one year of operation.
  }

  \label{fig:DICS_flux}
\end{figure}

In this appendix, we illustrate the flux of dark photons as a function of the
laboratory scattering angle $\phi$, considering a dark
photon mass of $m_{A^\prime}=1~\mathrm{eV}$ and a kinetic mixing parameter
$\varepsilon=10^{-7}$.

Figure~\ref{fig:DICS_flux} shows that dark-photon production is strongly
concentrated in the forward region for the benchmark beam and laser
configuration. Although the $A'$ itself is not detected, its angular
distribution provides useful information on the kinematics of the DICS process
and on the phase-space region in which the corresponding visible recoil-electron
signature is produced.
The two experimental strategies considered in this work exploit this invisible
final state in complementary ways. In the single-photon-veto strategy, the
reconstructed recoil-electron kinematics are used to identify events for which
an ordinary ICS photon would be expected inside the fiducial photon-detector
acceptance; the absence of such a photon constitutes the signal signature. In
the recoil-electron strategy, the DICS contribution is searched for through a
distortion of the reconstructed electron energy spectrum, with angular
information considered as an extension of the analysis.

The angular distribution shown in Figure~\ref{fig:DICS_flux} therefore provides
a useful illustration of the forward kinematics of the signal and complements
the corresponding ICS distributions discussed in the main text. In the
low-mass regime represented by the benchmark $m_{A'}=1~\mathrm{eV}$, the
normalized DICS kinematics are already close to their quasi-massless limit,
and hence to those of ordinary ICS. The sensitivity in this regime should
therefore not be interpreted as arising from angular separation between signal
and background, but from the missing-photon signature described in
Section~\ref{sec:veto-strategy}.

\acknowledgments

The authors thank Harry Westfahl from the Sirius accelerator for providing useful information about the accelerator, which is essential to developing this work. We also appreciate the discussions with Bertrand Laforge, Paolo Crivelli, Felix Kahlhoefer, Manfred Lindner and Seokhoon Yun to improve this proposal.  FSQ is supported by Simons Foundation (Award Number: 1023171-RC), Humboldt Foundation, and ANID - Millennium Science Initiative Program - ICN2019\textunderscore044., IIF-FINEP grant 213/2024, FAPESP Grants 2021/14335-0 and 2021/00449-4, CNPq grants 307130/2021-5, 403521/2024-6, 309465/2026-5, and 446711/2025-0. LA acknowledges support from the Coordenação de Aperfeiçoamento de Pessoal de Nível Superior (CAPES) under grants 88887.827404/2023-00 and 88881.128190/ 2025-01. LA also thanks the Department of Physics at the University of Oslo for support and useful discussions during the final stage of this work . GG acknowledges support from Universidad Nacional de Ingenieria funding Grant FC-A-PFR-06-2026. V.K. recognizes support by project SUMMIT BG-RRP-2.004-0008-C01 and by BNSF KP-06-COST/25 from 16.12.2024 based upon work from COST Action COSMIC WISPers CA21106 supported by COST (European Cooperation in Science and Technology). ASJ acknowledges support from FAPESP under the grant 2023/13126-4. JS and JDS acknowledge support from the UK Research and Innovation Future
Leader Fellowship MR/Y018656/1. We also thank IIP for the local cluster {\it bulletcluster}, which was instrumental in this work. 

\paragraph{Language and Editing Assistance}
The authors used Grammarly for assistance with language and editing. All scientific content and numerical results were independently developed and verified by the authors.


\bibliographystyle{JHEP}
\bibliography{biblio.bib}

\end{document}